# On the Feedback Capacity of Power Constrained Gaussian Noise Channels with Memory

Shaohua Yang, *Member, IEEE,* Aleksandar Kavčić, *Senior Member, IEEE*, and Sekhar Tatikonda, *Member, IEEE*

*Abstract*— For a stationary additive Gaussian-noise channel with a rational noise power spectrum of a finite-order $L$, we derive two new results for the feedback capacity under an average channel input power constraint. First, we show that a very simple feedback-dependent Gauss-Markov source achieves the feedback capacity, and that Kalman-Bucy filtering is optimal for processing the feedback. Based on these results, we develop a new method for optimizing the channel inputs for achieving the Cover-Pombra block-length-$n$ feedback capacity by using a dynamic programming approach that decomposes the computation into $n$ sequentially identical optimization problems where each stage involves optimizing $O(L^2)$ variables. Second, we derive the explicit maximal information rate for stationary feedback-dependent sources. In general, evaluating the maximal information rate for stationary sources requires solving only a few equations by simple non-linear programming. For first-order autoregressive and/or moving average (ARMA) noise channels, this optimization admits a closed form maximal information rate formula. The maximal information rate for stationary sources is a lower bound on the feedback capacity, and it equals the feedback capacity if the long-standing conjecture, that stationary sources achieve the feedback capacity, holds.

*Index Terms*— channel capacity, directed information, dynamic programming, feedback capacity, Gauss-Markov source, information rate, intersymbol interference, Kalman-Bucy filter, linear Gaussian noise channel, noise whitening filter

## I. INTRODUCTION

We consider discrete-time power-constrained linear Gaussian noise channels, where the signal is corrupted by an additive Gaussian random process. When the channel is memoryless, i.e., when the channel is corrupted by additive *white* Gaussian noise (AWGN), Shannon [1] provided a simple formula for computing the feed-forward channel capacity, and he also proved that feedback does not increase the channel capacity [2].

For Gaussian noise channels with memory, i.e., when the channel noise is correlated, the feed-forward channel capacity can be determined by the power-spectral-density water-filling method [3], [4], [5], [6]. However, with noiseless feedback, i.e., when the transmitter knows without error all previous channel outputs, it has been a long-standing open problem

Manuscript submitted on October 22, 2003; first revision December 10, 2004; second revision December 23, 2005; third revision June 10, 2006; fourth revision September 25, 2006.
This work was supported by the National Science Foundation under Grant No. CCR-9904458 and by the National Storage Industry Consortium.
Shaohua Yang is with the Marvell Semiconductor Inc, Santa Clara, CA 95054. Email: syang@marvell.com. Aleksandar Kavčić is with the University of Hawaii, Honolulu, HI 96822. Sekhar Tatikonda is with the Department of Electrical Engineering, Yale University, New Haven, CT 06520. Email: sekhar.tatikonda@yale.edu.

to explicitly characterize the optimal (or feedback-capacity-achieving) signal and thus compute the feedback channel capacity. Cover and Pombra [7] defined the $n$-block feedback capacity and formulated the $n$-block feedback capacity computation as an optimization problem. Upper and lower bounds on the feedback capacity were established. For example, it is known that the feedback capacity can never exceed the capacity without feedback (feed-forward capacity) by more than 0.5 bit/channel-use [7], or that the feedback capacity can never be more than double the feed-forward capacity [8], [9]. Somewhat tighter upper bounds can be computed by numerical techniques as in [10]. Butman [11] devised a feedback code that recursively transmits the message through the channel, which achieves a higher rate than the feed-forward channel capacity for certain linear Gaussian channels. For first-order autoregressive (AR) noise channels, the closed-form information rate obtained by Butman [11] is very close to the tightest upper bound and thus has been conjectured to be the real feedback capacity, but a rigorous proof has been missing. Ordentlich [12] characterized an optimal feedback coding scheme for moving-average Gaussian noise channels. Tatikonda [13] formulated the feedback channel capacity in terms of the directed information rate. Ihara [14] studied the continuous-time Gaussian noise channel with feedback.

There are two main contributions in this work:

1) We characterize the optimal signaling and feedback strategy for achieving the $n$-block feedback capacity of a power constrained linear Gaussian channel with a rational power spectrum. We show that the optimal source is a simple feedback-dependent Gauss-Markov source and that a Kalman-Bucy filter is optimal for processing the feedback. This leads to a reformulation of the problem as a stochastic control optimization problem. As a result, a new method based on dynamic programming is derived to optimize the source and thus compute the $n$-block feedback capacity. For computing the $n$-block feedback capacity, the new method decomposes the computation into $n$ identical sequential optimization problems with each stage involving only $O(L^2)$ variables, where $L$ is the order of the rational power spectral density of the noise (or channel).

We prove that a Gauss Markov source (channel input process) $X_t$ of the following form (also depicted in Fig. 1) is optimal

$$X_t = \underline{d}_t^{\mathrm{T}} \underbrace{(\underline{S}_{t-1} - \underline{m}_{t-1})}_{\text{Kalman innovation}} + e_t Z_t,$$

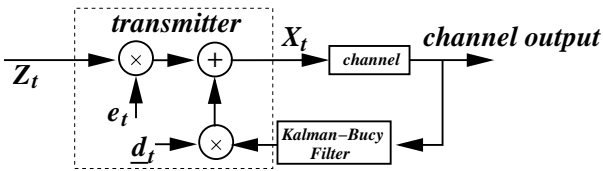

Fig. 1. Optimal source for achieving the feedback capacity

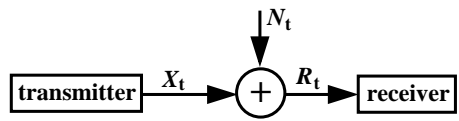

Fig. 2. A discrete-time linear Gaussian noise channel

where $\underline{d}_t$ (a vector of dimension $L$) and $e_t$ are predetermined coefficients, vectors $\underline{S}_{t-1}$ and $\underline{m}_{t-1}$ are the channel state (of dimension $L$) and the channel state estimate (computed by the Kalman-Bucy filter in the feedback loop), respectively, and $Z_t$ is a zero-mean unit-variance Gaussian random variable that is independent of prior channel inputs, outputs and noise.

We note that the Kalman-Bucy filter developed in this paper estimates the channel intersymbol-interference state $S_t$ and takes a different form from the filters used in Schalkwijk [15] or Butman [11] which recursively estimate the transmitted *message*, but it is closely related. The optimal channel input derived in this paper includes a recursive estimate term (the Kalman innovation) and a novelty term ($e_t Z_t$), thus it could be equivalent to the recursive message estimating and transmitting scheme in [15], [11] only if the optimal value of the novelty term $e_t Z_t$ is zero, which is an open problem that we leave unresolved.

2) We derive an explicit formula for the maximal information rate achieved by (asymptotically) stationary feedback-dependent sources. This represents a lower bound on the feedback capacity. We note that it is a long-standing conjecture that a stationary source achieves the feedback capacity, and this conjecture is not proved in this paper and is still open. If the optimal Kalman-Bucy filter for processing the feedback (optimized over both stationary and non-stationary feedback-dependent sources) has a steady state, i.e., if the optimal Kalman-Bucy filter becomes stationary as $n \to \infty$, the feedback channel capacity exists and equals the maximal information rate for stationary sources.

For the case of the first order autoregressive (AR) noise channels, our optimal signaling scheme for achieving the maximal stationary-source information rate turns out to be the same as Butman's code [11].

**Paper organization:** We introduce the Guassian noise channel model in Section II. For convenience the problem is reformulated in the state-space (or state-machine) realization context. In Section III, the $n$-block feedback capacity is expressed in a form that is suitable for solving the optimization problem using dynamic programming techniques. In Section IV, we show that Gauss-Markov sources achieve the $n$-block feedback capacity and that a Kalman-Bucy filter is optimal for processing the feedback. Section V is devoted to solving the feedback capacity computation problem. A simple feedback-capacity-achieving signaling scheme is explicitly characterized and a dynamic programming algorithm to optimize the source is presented. In Section VI, we derive an explicit formula for the maximal feedback information rate achieved by stationary sources, which represents a lower bound on the feedback capacity. This maximal feedback information rate can be evaluated by nonlinear programming techniques. We solve the nonlinear programming problem in closed form for first-order autoregressive/moving-average (ARMA) channels. Section VII concludes the paper.

## II. POWER-CONSTRAINED LINEAR GAUSSIAN NOISE CHANNEL MODEL

Let $t \in \mathbb{Z}$ denote the discrete time index, and let the random variable $X_t$ denote the channel input at time $t$. As depicted in Figure 2, additive stationary Gaussian noise $N_t$ corrupts the channel input $X_t$ to form the channel output random variable

$$R_t = X_t + N_t. \quad (1)$$

It is assumed that the power spectral density function of the noise process $N_t$ is known and is denoted as $S_N(\omega)$. In its most general formulation, the power spectral density function of the Gaussian noise process can be an arbitrary nonnegative function defined on the interval $\omega \in (-\pi, \pi]$, such that it is even, i.e., $S_N(\omega) = S_N(-\omega)$, and its power is finite

$$\sigma_N^2 = \frac{1}{2\pi} \int_{-\pi}^{\pi} S_N(\omega) d\omega < \infty. \quad (2)$$

Further, it is required that the average signal power be bounded by a known value $P$ from above, i.e., we have an average channel-input power constraint

$$\lim_{n \to \infty} \mathrm{E}\left[\frac{1}{n} \sum_{t=1}^{n} (X_t)^2\right] \leq P. \quad (3)$$

If a finite block length $n$ is considered as in [7], then the power constraint becomes

$$\mathrm{E}\left[\frac{1}{n} \sum_{t=1}^{n} (X_t)^2\right] \leq P. \quad (4)$$

The Gaussian noise channel has memory when the noise is correlated. A correlated noise process $N_t$ can be obtained by passing a white Gaussian noise process $W_t$ through a linear filter defined by a constant-coefficient difference equation of the following form

$$N_t = W_t - \sum_{l=1}^{L} a_l W_{t-l} - \sum_{l=1}^{L} c_l N_{t-l}, \quad (5)$$

where $W_t$ is a white Gaussian random process with variance $\sigma_W^2$. It is easy to verify that the $z$-domain transfer

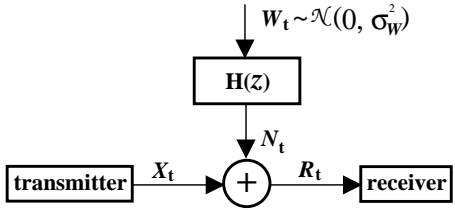

Fig. 3. A linear Gaussian noise channel with a rational noise filter

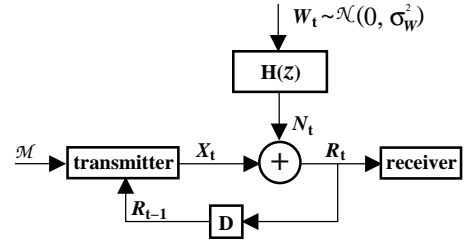

Fig. 4. Linear Gaussian noise channel with feedback.

function of the linear noise filter in (5) is

$$H(z) = \frac{1 - \sum\limits_{l=1}^{L} a_l z^{-l}}{1 + \sum\limits_{l=1}^{L} c_l z^{-l}}, \quad (6)$$

and that the power spectral density function of the noise process $N_t$ is rational, i.e.,

$$S_N(\omega) = \sigma_W^2 \left| H(e^{j\omega}) \right|^2$$
$$= \sigma_W^2 \frac{\left(1 - \sum\limits_{l=1}^{L} a_l e^{-jl\omega}\right)\left(1 - \sum\limits_{l=1}^{L} a_l e^{jl\omega}\right)}{\left(1 + \sum\limits_{l=1}^{L} c_l e^{-jl\omega}\right)\left(1 + \sum\limits_{l=1}^{L} c_l e^{jl\omega}\right)}. \quad (7)$$

Such a *linear Gaussian noise channel* with memory is depicted in Figure 3.

It is known that any power spectral density can be approximated arbitrarily closely by the rational function (7) if the model size (memory length) parameter $L$ is chosen to be large enough [17]. So, restriction (5) is not strong, but as we will see, it is very much needed in the subsequent analysis. Further, without loss of generality, we may assume that the function $H(z)$ has all its poles and zeros inside the unit circle and that no poles or zeros occur at the origin $z = 0$. Thus, $H(z)$ is a causal and stable minimum phase linear filter [17], and its inverse is also a causal and stable minimum phase linear filter.

For the linear Gaussian noise channel, the channel capacity [1] is defined as the maximal amount of information that could be transmitted per channel use with an arbitrarily small error probability. It is well known that noiseless channel output feedback, i.e., error-free observation of the channel output by the transmitter, may increase the channel capacity [11]. Thus, if we denote the feed-forward channel capacity as $C$ and the feedback channel capacity as $C^{\text{fb}}$, then $C \leq C^{\text{fb}}$.

As illustrated in Figure 4, the transmitter knows at time $t$, without error, all previous realizations of the channel outputs before it forms and sends out the next channel input signal. The transmission starts at time $t = 1$. We also assume that prior to time $t = 1$, a known signal is transmitted, e.g., $X_t = 0$ for $t \leq 0$, and thus both the transmitter and receiver know the prior channel outputs $R_{-\infty}^0 = N_{-\infty}^0 = n_{-\infty}^0$. This is equivalent to assuming that both the receiver and the transmitter know $N_{-\infty}^0 = n_{-\infty}^0$ and $W_{-\infty}^0 = w_{-\infty}^0$. This assumption on the channel inputs and outputs prior to transmission is mainly for simplifying the analysis in the state-space setting, which we formally introduce in Section II-B.

We denote the message as $\mathcal{M}$ and encode it into $n$ feedback-dependent channel input signal variables $X_1^n$ by using a feedback encoder, which in its most general form is

$$X_t = X_t\left(\mathcal{M}, R_1^{t-1}, n_{-\infty}^0\right). \quad (8)$$

The receiver tries to decode the message $\mathcal{M}$ based on the realization of the channel output variables $R_1^n = r_1^n$.

We next summarize some of the relevant known results.

### A. Butman's Recursive Feedback Coding Scheme

Butman [11] considered a very general communication system for colored Gaussian noise channels with or without feedback. The feedback considered can be noiseless or noisy. If we omit the noisy feedback scenario, the transmitted signal $X_t$ is chosen as a linear combination of the feedback and the Gaussian message $\mathcal{M} = \theta$ as

$$X_t = \delta_t \theta + \sum_{i=1}^{t-1} a_{ti} R_i. \quad (9)$$

With this transmission model, Butman optimized the signaling for certain Gaussian noise channels with feedback. For first-order autoregressive noise channels, the following feedback scheme achieves the maximal information rate under the model in (9). At each time instant $t > 0$, the transmitter computes the receiver-side minimum-variance estimate of the message

$$\tilde{\theta} = \mathrm{E}\left[\theta \,\middle|\, R_1^{t-1} = r_1^{t-1}\right] \quad (10)$$

and transmits

$$X_t = \delta_t\left(\theta - \tilde{\theta}\right). \quad (11)$$

By optimizing the coefficients $\delta_t$ to maximize the information rate subject to the power constraint, a tight lower bound on the feedback capacity was obtained, which was shown to be greater than the feed-forward channel capacity. This result was generalized to higher order channels [18]. Proving that Butman's coding scheme (9) achieves the capacity is still an open problem.

### B. $n$-Block Feedback Capacity

Cover and Pombra [7] introduced the (block-length-$n$) feedback capacity as the maximal achievable information rate (i.e., maximal information per channel use) for finite block length (or finite horizon) $n$, and showed that a sequence of codes exists that can achieve the capacity as $n \to \infty$.

Since we are assuming that the noise realization $n_{-\infty}^0$ is known both to the receiver and the transmitter prior to the start of transmission, we can express the Cover-Pombra $n$-block feedback capacity as $C^{\text{fb}(n)} \triangleq \max \frac{1}{n} I\left(\mathcal{M}; R_1^n \,\big|\, n_{-\infty}^0\right)$, where the maximization is taken under a finite-horizon power constraint

$$\mathrm{E}\left[\frac{1}{n}\sum_{t=1}^n (X_t)^2 \,\bigg|\, N_{-\infty}^0 = n_{-\infty}^0\right] \leq P. \quad (12)$$

Define the following covariance matrices

$$\begin{aligned}
\mathbf{K}_N^{(n)} &\triangleq \mathrm{E}\left[N_1^n \cdot (N_1^n)^{\mathrm{T}} \,\big|\, N_{-\infty}^0 = n_{-\infty}^0\right] \\
\mathbf{K}_X^{(n)} &\triangleq \mathrm{E}\left[X_1^n \cdot (X_1^n)^{\mathrm{T}} \,\big|\, X_{-\infty}^0 = n_{-\infty}^0\right] \\
\mathbf{K}_{X+N}^{(n)} &\triangleq \mathrm{E}\left[(X_1^n + N_1^n) \cdot (X_1^n + N_1^n)^{\mathrm{T}} \,\big|\, N_{-\infty}^0 = n_{-\infty}^0\right].
\end{aligned}$$

As shown in [7], the $n$-block feedback capacity (or maximal information rate) equals

$$C^{\text{fb}(n)} = \max_{\frac{1}{n}\mathrm{tr}\left(\mathbf{K}_X^{(n)}\right) \leq P} \frac{1}{2n} \log \frac{\left|\mathbf{K}_{X+N}^{(n)}\right|}{\left|\mathbf{K}_N^{(n)}\right|}. \quad (13)$$

Here, the maximization is taken over all channel input variables $X_1^n$, which take the following linear form

$$X_t = \sum_{i=1}^{t-1} b_{ti} N_i + V_t, \text{ for } t = 1, 2, \ldots, n, \quad (14)$$

where $b_{ti}$ are the coefficients that need to be optimized, and the random variables $V_t$ have a Gaussian distribution whose covariance matrix needs to be optimized.

The $n$-block feedback capacity computation problem in (13) is formulated for an arbitrary noise covariance matrix $\mathbf{K}_N^{(n)}$. This optimization problem can be solved by methods given in [19]. The number of unknown variables in (14) is $O(n^2)$. For the feedback code in (14) the transmitter would need to remember and utilize all previous channel output realizations $R_1^{t-1} = r_1^{t-1}$ (or the noise realizations $N_1^{t-1} = n_1^{t-1}$) so as to form the next input signal $X_t = x_t$ by using the linear equation (14). Thus, the encoding complexity grows with time $t$ in general.

In the following, we reformulate the linear Gaussian noise channel in Figure 4 as a state-space channel model. The state-space formalism permits a very simple feedback-capacity-achieving signaling scheme whose encoding complexity is constant for any time instant $t \geq 1$. For this optimal signaling scheme we show in this paper, the number of variables grows as $O(n)$, and the encoding complexity is fixed for all $t > 1$.

### C. An Equivalent State-Space Gaussian Noise Channel Model

The filter $H(z)$ in (6) is modeled as a rational filter with all poles and zeros inside the unit circle and no poles or zeros in the origin. The inverse of such a filter exists and is also invertible, so we may filter (without causing latency) the channel output by a noise whitening filter $H^{-1}(z)$ to get an equivalent state-space (intersymbol interference) channel model with white noise (depicted in Figure 5)

$$Y(z) = H^{-1}(z)R(z) = H^{-1}(z)X(z) + W(z), \quad (15)$$

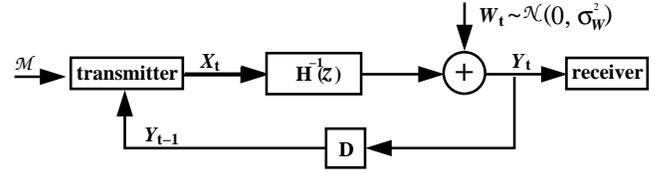

Fig. 5. An equivalent linear Gaussian noise channel model

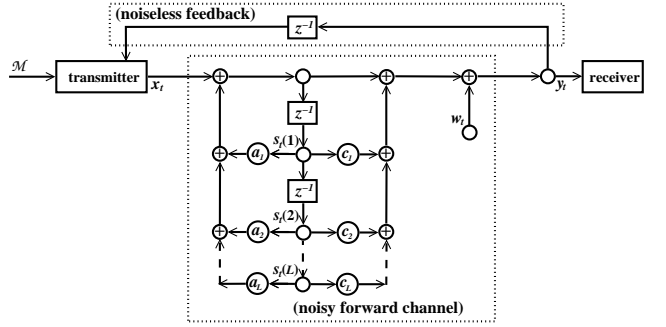

Fig. 6. An $L^{th}$-order LTI Gaussian noise channel with noiseless feedback

or equivalently

$$Y_t - \sum_{l=1}^L a_l Y_{t-l} = X_t + \sum_{l=1}^L c_l X_{t-l} + W_t - \sum_{l=1}^L a_l W_{t-l}, \quad (16)$$

where $W_t$ is an independent and identically distributed (i.i.d.) Gaussian random process with variance $\sigma_W^2$.

The channel model (16), depicted in Figure 5, is a channel model with intersymbol interference due to the filter $H^{-1}(z)$. Note that the transmitter obtains $Y_t$ by filtering the original channel output $R_t$ using the filter $H^{-1}(z)$. The filters $H(z)$ and $H^{-1}(z)$ are causal, minimum phase and invertible, and thus do not cause any delay, see [17]. Thus, the two channel models depicted in Figure 4 and Figure 5 could be converted into each other's form by filtering their channel outputs using rational causal delay-free filters $H(z)$ and $H^{-1}(z)$, respectively. Hence, the two channel models are mathematically equivalent and have the same feedback (or feed-forward) channel capacities.

The rational filter $H^{-1}(z)$ can be realized by shift registers [17], and the channel model depicted in Figure 5 can thus be represented as a state-space (or state-machine) channel model with noiseless feedback. As depicted in Figure 6, we only need to consider a linear time-invariant (LTI) filter [17] observed through an additive white Gaussian noise (AWGN) channel, see equation (15). The LTI filter in the channel is completely characterized by its order $L$ and the two vectors of tap coefficients

$$\underline{a} \triangleq [a_1, a_2, \cdots, a_L]^{\mathrm{T}}, \quad (17)$$
$$\underline{c} \triangleq [c_1, c_2, \cdots, c_L]^{\mathrm{T}}, \quad (18)$$

which are also coefficients of the rational filter $H(z)$ in equation (6).

Let $X_t$ be the channel input at time $t$ whose realization is denoted by $x_t$. Let $Y_t$ be the noisy channel output at time $t$ whose realization is denoted by $y_t$ (which is the *filtered* channel output $R_t$ of the linear Gaussian noise channel in Figure 4). Let the vector of values stored in the shift registers of the LTI filter, i.e., $\underline{S}_t \triangleq [S_t(1), S_t(2), \ldots, S_t(L)]^\mathrm{T}$, be the channel *state* vector, and denote the state realization by $\underline{s}_t \triangleq [s_t(1), s_t(2), \ldots, s_t(L)]^\mathrm{T}$. Since it is assumed that the channel inputs $X_{-\infty}^0$ are known prior to the start of transmission at time $t = 1$, the state realization $\underline{S}_0 = \underline{s}_0$ is also known. Then, the channel can be described by the following assumptions:

1) The forward channel satisfies a state-space model, i.e.,

$$\underline{S}_t = \mathbf{A}\underline{S}_{t-1} + \underline{b}X_t \tag{19}$$
$$Y_t = (\underline{a} + \underline{c})^\mathrm{T} \underline{S}_{t-1} + X_t + W_t, \tag{20}$$

where $W_t$ is white Gaussian noise with variance $\sigma_W^2$. The square matrix $\mathbf{A}$ and the vector $\underline{b}$ are time-invariant as follows

$$\mathbf{A} \triangleq \begin{bmatrix} a_1 & a_2 & \ldots & a_{L-1} & a_L \\ 1 & 0 & \ldots & 0 & 0 \\ 0 & 1 & \ldots & 0 & 0 \\ \vdots & \vdots & \ddots & \vdots & \vdots \\ 0 & 0 & \ldots & 1 & 0 \end{bmatrix}, \quad \underline{b} \triangleq \begin{bmatrix} 1 \\ 0 \\ 0 \\ \vdots \\ 0 \end{bmatrix}. \tag{21}$$

The LTI filter described by the coefficient vectors $\underline{a}$ and $\underline{c}$ is invertible and stable. The initial state $\underline{S}_0 = \underline{s}_0$ is known to both the encoder and the decoder.

2) The feedback link is noiseless and causal, i.e., the encoder, before sending out symbol $X_t$, knows without error all previous channel outputs $Y_1^{t-1} = y_1^{t-1}$.

3) The average channel input power is constrained by

$$\frac{1}{n} \sum_{t=1}^{n} \mathrm{E}\left[(X_t)^2 \,|\, \underline{S}_0 = \underline{s}_0\right] \leq P, \tag{22}$$

where $n$ is the total number of input symbols $X_1^n$ that are used to encode the message $\mathcal{M}$.

From assumptions 1)-3), we have the following:

I) For any given initial channel state $\underline{S}_0 = \underline{s}_0$, the sequences $\underline{S}_1^t$ and $X_1^t$ determine each other uniquely because of the linear equation (19). If a state sequence does not conform to the channel law (19), it is called *invalid*. We only need to consider *valid* state sequences throughout. From the definition of the state $S_t$ in (19), we see that when the initial state $s_0$ is known, the channel input sequence $X_t$ and the channel state sequence $S_t$ are in a 1-to-1 relationship. So, the mutual information rate between the channel input sequence (source) and the channel output sequence is the same as the mutual information rate between the channel state sequence and the channel output sequence. Thus, it is valid to refer to the state sequence as the *source* sequence.

II) Given the channel state pair $\underline{S}_{t-1}^t = (\underline{S}_{t-1}, \underline{S}_t)$, the channel output $Y_t$ is statistically independent of previous channel states $\underline{S}_0^{t-2}$ and outputs $Y_1^{t-1}$, that is

$$P_{Y_t | \underline{S}_0^t, Y_1^{t-1}}\left(y_t \,|\, \underline{s}_0^t, y_1^{t-1}\right) = P_{Y_t | \underline{S}_{t-1}^t}\left(y_t \,|\, \underline{s}_{t-1}^t\right). \tag{23}$$

Since the variance of the white Gaussian noise $W_t$ is $\sigma_W^2$, from equation (20) we have the following conditional differential entropy of the channel output

$$\begin{aligned} h\left(Y_t \,|\, \underline{S}_0^t, Y_1^{t-1}\right) &= h\left(Y_t \,|\, \underline{S}_{t-1}^t\right) \\ &= h(W_t) \\ &= \frac{1}{2}\log(2\pi\mathrm{e}\sigma_W^2). \end{aligned} \tag{24}$$

III) Since the sequences $X_t$ and $S_t$ uniquely determine each other for any given value of $s_0$, we can characterize the source distribution either in terms of channel inputs $X_t$ as

$$P_t\left(x_t | \underline{s}_0^{t-1}, y_1^{t-1}\right) \triangleq P_{X_t | \underline{S}_0^{t-1}, Y_1^{t-1}}\left(x_t | \underline{s}_0^{t-1}, y_1^{t-1}\right), \tag{25}$$

or in terms of channel states $S_t$ as

$$P_t\left(\underline{s}_t | \underline{s}_0^{t-1}, y_1^{t-1}\right) \triangleq P_{\underline{S}_t | \underline{S}_0^{t-1}, Y_1^{t-1}}\left(\underline{s}_t | \underline{s}_0^{t-1}, y_1^{t-1}\right). \tag{26}$$

For the Gaussian noise channel formulated in the state-space framework, the feedback capacity for a finite horizon $n$ equals [7]

$$\begin{aligned} C^{\mathrm{fb}(n)} &= \max \frac{1}{n} I\left(\mathcal{M}; Y_1^n \,|\, \underline{S}_0 = \underline{s}_0\right) \\ &= \max \frac{1}{n} \left[h\left(Y_1^n \,|\, \underline{S}_0 = \underline{s}_0\right) - h\left(W_1^n\right)\right], \end{aligned} \tag{27}$$

where the maximization is over the channel input distribution (26) and is subject to the average input power constraint (22). In the subsequent sections, we focus on characterizing the optimal signaling and the feedback capacity for the state-space channel model depicted in Figure 6.

## III. $n$-BLOCK FEEDBACK CAPACITY

In this section, we manipulate the mutual information $I\left(\mathcal{M}; Y_1^n \,|\, \underline{S}_0 = \underline{s}_0\right)$ into a form that is suitable for evaluation and analysis. Let the source distribution induced by the encoder $X_t = X_t\left(\mathcal{M}, Y_1^{t-1}\right)$, i.e., the *set* of all valid conditional probability density functions defined in (25) or (26), be denoted by

$$\mathcal{P} \triangleq \{P_t\left(\underline{s}_t | \underline{s}_0^{t-1}, y_1^{t-1}\right), t = 1, 2, \ldots\}. \tag{28}$$

The $n$-block feedback capacity computation problem (13) becomes (see [7])

$$\begin{aligned} C^{\mathrm{fb}(n)} &= \max_{\mathcal{P}} \frac{1}{n} I\left(\mathcal{M}; Y_1^n \,|\, \underline{S}_0 = \underline{s}_0\right) \\ &= \max_{\mathcal{P}} \frac{1}{n} \left[h\left(Y_1^n \,|\, \underline{S}_0 = \underline{s}_0\right) - h\left(W_1^n\right)\right], \end{aligned} \tag{29}$$

where the maximization is over the source distribution $\mathcal{P}$. Note that in (28) we temporarily ignore the linear structure of the capacity-achieving signaling (14) derived by Cover and Pombra in [7], and consider the distribution $\mathcal{P}$ instead because we first want to show that the a Markov structure of the distribution $\mathcal{P}$ is sufficient. Consequently, we will show that a linear signaling scheme, taking a very simple form, is sufficient for achieving the $n$-block capacity.

The differential entropy of the channel noise in (29) can be alternatively expressed as

$$h(W_1^n) = \sum_{t=1}^{n} h(W_t) = \sum_{t=1}^{n} h\left(Y_t \,\middle|\, Y_1^{t-1}, \underline{S}_1^t, \underline{S}_0 = \underline{s}_0\right). \quad (30)$$

Here, the equalities follow from the fact that the whitened noise sequence $W_t$ is independent and identically distributed (i.i.d.) and from the channel assumptions in Section II-C, see (24). By substituting (30) into the expression for $I(\mathcal{M}; Y_1^n \,|\, \underline{S}_0 = \underline{s}_0)$ in (29), we arrive at several equivalent expressions for $I(\mathcal{M}; Y_1^n \,|\, \underline{S}_0 = \underline{s}_0)$

$$\begin{aligned}
& I\left(\mathcal{M}; Y_1^n \,\middle|\, \underline{S}_0 = \underline{s}_0\right) \\
&= \sum_{t=1}^{n} \left[h\left(Y_t \,\middle|\, Y_1^{t-1}, \underline{s}_0\right) - \frac{1}{2}\log\left(2\pi e \sigma_W^2\right)\right] \quad (31) \\
&= \sum_{t=1}^{n} \left[h\left(Y_t \,\middle|\, Y_1^{t-1}, \underline{s}_0\right) - h\left(Y_t \,\middle|\, Y_1^{t-1}, \underline{S}_1^t, \underline{s}_0\right)\right] \quad (32) \\
&= \sum_{t=1}^{n} \left[h\left(Y_t \,\middle|\, Y_1^{t-1}, \underline{s}_0\right) - h\left(Y_t \,\middle|\, Y_1^{t-1}, \underline{S}_{t-1}^t, \underline{s}_0\right)\right] \quad (33) \\
&= \sum_{t=1}^{n} I\left(\underline{S}_{t-1}^t; Y_t \,\middle|\, Y_1^{t-1}, \underline{s}_0\right), \quad (34)
\end{aligned}$$

where equalities in (31) and (32)[1] follow from the definition of mutual information; the equality in (33) comes from the chain rule for mutual information and the channel assumptions, see (24); and equality (34) follows from the definition of mutual information. The terms in the sums (31) through (34) represent the amount of information that every channel use contributes to the total transmitted information.

Now, the feedback capacity can be expressed as

$$\begin{aligned}
C^{\text{fb}(n)} &= \max_{\mathcal{P}} \frac{1}{n} I\left(\mathcal{M}; Y_1^n \,\middle|\, \underline{S}_0 = \underline{s}_0\right) \\
&= \max_{\mathcal{P}} \frac{1}{n} \sum_{t=1}^{n} I\left(\underline{S}_{t-1}^t; Y_t \,\middle|\, Y_1^{t-1}, \underline{s}_0\right), \quad (35)
\end{aligned}$$

where the maximization is taken over the set $\mathcal{P}$ of valid feedback-dependent source distribution functions (28), see [7] or [13] for the proof.

A source distribution $\mathcal{P}$ is called *optimal* if it maximizes $I(\mathcal{M}; Y_1^n \,|\, \underline{S}_0 = \underline{s}_0)$ and thus achieves the $n$-block feedback capacity in (35). Since the information rate $\frac{1}{n} I(\mathcal{M}; Y_1^n \,|\, \underline{S}_0 = \underline{s}_0)$ is linearly proportional to the entropy rate of the channel output, see (31), a feedback-dependent source is optimal if and only if it maximizes the entropy rate of the channel output process $Y_t$. For a linear Gaussian noise channel, the feedback capacity is achieved by a Gaussian source distribution, see [7], [13]. Therefore, in the sequel, without loss of optimality, we only consider feedback-dependent *Gaussian* source distributions.

Since the number of arguments for the probability density function (pdf) $P_t(\underline{s}_t \,|\, \underline{s}_0^{t-1}, y_1^{t-1})$ increases linearly with time $t$, it is hard to directly find the optimal distribution $\mathcal{P}$ in (28)

---
[1]The right-hand side of equation (32) is defined by Marko [20] and Massey [21] as the directed information from the channel state to the channel output.

and thereby compute the $n$-block feedback capacity $C^{\text{fb}(n)}$ for large block length $n$. However, by working on a state-space channel model as defined in Section II-C, we are able to significantly simplify the problem and derive a simple dynamic programming method to compute the $n$-block feedback capacity.

## IV. $n$-BLOCK FEEDBACK-CAPACITY-ACHIEVING STRATEGY

### A. Gauss-Markov Sources Achieve the Feedback Capacity

In the following analysis, we note that since the initial channel state $\underline{s}_0$ is known according to the channel assumption in Section II-C, for notational simplicity, we will not explicitly write the dependence on $\underline{s}_0$ when obvious.

*Theorem 1:* [*Feedback-dependent Gauss-Markov sources achieve the feedback capacity*] For the power constrained Gaussian channel, a feedback-dependent Gauss-Markov source distribution (not necessarily stationary) of the following form

$$\mathcal{P}^{\text{GM}} \triangleq \left\{P_t\left(\underline{s}_t \,\middle|\, \underline{s}_{t-1}, y_1^{t-1}\right), t = 1, 2, \ldots\right\} \quad (36)$$

achieves the $n$-block feedback capacity $C^{\text{fb}(n)}$, for any block length $n$.

*Proof:* We adopt the following proof strategy. As shown in [7], we only need to consider feedback-dependent Gaussian source distributions. We take an arbitrary feedback-dependent Gaussian (not necessarily Markov) source distribution $\mathcal{P}_1$ of the form in (28). From the source $\mathcal{P}_1$, we obtain the marginal state transition probabilities, from which we construct a feedback-dependent Gauss-*Markov* source $\mathcal{P}_2$ of the form in (36). We then show that sources $\mathcal{P}_1$ and $\mathcal{P}_2$, though different in general, induce the same information $I(\mathcal{M}; Y_1^n \,|\, \underline{S}_0 = \underline{s}_0)$. Using this argument, we will show that for any optimal feedback-dependent Gaussian source distribution that achieves the $n$-block feedback capacity, there exists a feedback-dependent Gauss-Markov source distribution (not necessarily stationary) that also achieves the $n$-block feedback capacity.

Let $\mathcal{P}_1$ be any valid feedback-dependent Gaussian source distribution defined as

$$\mathcal{P}_1 \triangleq \left\{P_t\left(\underline{s}_t \,\middle|\, \underline{s}_0^{t-1}, y_1^{t-1}\right), t = 1, 2, \cdots\right\}. \quad (37)$$

From the source $\mathcal{P}_1$, we define the sequence of conditional marginal pdf's $Q_t(\underline{s}_t \,|\, \underline{s}_{t-1}, y_1^{t-1})$, for $t = 1, 2, \cdots$, as follows

$$Q_t\left(\underline{s}_t \,\middle|\, \underline{s}_{t-1}, y_1^{t-1}\right) \triangleq P_{\underline{S}_t \,|\, \underline{S}_{t-1}, Y_1^{t-1}}^{(\mathcal{P}_1)}\left(\underline{s}_t \,\middle|\, \underline{s}_{t-1}, y_1^{t-1}\right) \quad (38)$$

$$= \frac{\int \left[\prod_{\tau=1}^{t-1} P_\tau\left(\underline{s}_\tau \,\middle|\, \underline{s}_0^{\tau-1}, y_1^{\tau-1}\right) f_{Y_\tau \,|\, \underline{S}_{\tau-1}^\tau}\left(y_\tau \,\middle|\, \underline{s}_{\tau-1}^\tau\right)\right] P_t\left(\underline{s}_t \,\middle|\, \underline{s}_0^{t-1}, y_1^{t-1}\right) d\underline{s}_1^{t-2}}{\int \left[\prod_{\tau=1}^{t-1} P_\tau\left(\underline{s}_\tau \,\middle|\, \underline{s}_0^{\tau-1}, y_1^{\tau-1}\right) f_{Y_\tau \,|\, \underline{S}_{\tau-1}^\tau}\left(y_\tau \,\middle|\, \underline{s}_{\tau-1}^\tau\right)\right] d\underline{s}_1^{t-2}}. \quad (39)$$

Using the functions $Q_t(\underline{s}_t \,|\, \underline{s}_{t-1}, y_1^{t-1})$, we construct a feedback-dependent Markov (not necessarily stationary) source distribution $\mathcal{P}_2$ as

$$\mathcal{P}_2 = \left\{Q_t\left(\underline{s}_t \,\middle|\, \underline{s}_{t-1}, y_1^{t-1}\right), t = 1, 2, \cdots\right\}. \quad (40)$$

The source $\mathcal{P}_2$ induces the following joint pdf of the channel states $\underline{S}_{t-1}^t$ and outputs $Y_1^t$

$$P_{\underline{S}_{t-1}^t, Y_1^t | \underline{S}_0}^{(\mathcal{P}_2)} \left( \underline{s}_{t-1}^t, y_1^t | \underline{s}_0 \right)$$

$$= \int P_{\underline{S}_1^t, Y_1^t | \underline{S}_0}^{(\mathcal{P}_2)} \left( \underline{s}_1^t, y_1^t | \underline{s}_0 \right) \mathrm{d}\underline{s}_1^{t-2} \quad (41)$$

$$= \int \prod_{\tau=1}^t Q_\tau \left( \underline{s}_\tau | \underline{s}_{\tau-1}, y_1^{\tau-1} \right) P_{Y_\tau | \underline{S}_{\tau-1}^\tau} \left( y_\tau | \underline{s}_{\tau-1}^\tau \right) \mathrm{d}\underline{s}_1^{t-2}. \quad (42)$$

We next show by induction that the joint distribution (42) of $\underline{S}_{t-1}^t$ and $Y_1^t$ induced by the source $\mathcal{P}_2$ is the same as the one induced by the source $\mathcal{P}_1$, i.e.,

$$P_{\underline{S}_{t-1}^t, Y_1^t | \underline{S}_0}^{(\mathcal{P}_2)} \left( \underline{s}_{t-1}^t, y_1^t | \underline{s}_0 \right) = P_{\underline{S}_{t-1}^t, Y_1^t | \underline{S}_0}^{(\mathcal{P}_1)} \left( \underline{s}_{t-1}^t, y_1^t | \underline{s}_0 \right) \quad (43)$$

$$= \int P_{\underline{S}_1^t, Y_1^t | \underline{S}_0}^{(\mathcal{P}_1)} \left( \underline{s}_1^t, y_1^t | \underline{s}_0 \right) \mathrm{d}\underline{s}_1^{t-2} \quad (44)$$

$$= \int \prod_{\tau=1}^t P_\tau \left( \underline{s}_\tau | \underline{s}_0^{\tau-1}, y_1^{\tau-1} \right) P_{Y_\tau | \underline{S}_{\tau-1}^\tau} \left( y_\tau | \underline{s}_{\tau-1}^\tau \right) \mathrm{d}\underline{s}_1^{t-2}. \quad (45)$$

For $t = 1$, by the definition (39) of source $\mathcal{P}_2$ we have

$$Q_1(\underline{s}_1 | \underline{s}_0) = P_1(\underline{s}_1 | \underline{s}_0). \quad (46)$$

We verify the to-be-proved equality (43) for $t=1$ by noting that

$$P_{\underline{S}_1, Y_1 | \underline{S}_0}^{(\mathcal{P}_2)} (\underline{s}_1, y_1 | \underline{s}_0) = Q_1(\underline{s}_1 | \underline{s}_0) P_{Y_1 | \underline{S}_0^1} (y_1 | \underline{s}_0^1) \quad (47)$$

$$= P_1(\underline{s}_1 | \underline{s}_0) P_{Y_1 | \underline{S}_0^1} (y_1 | \underline{s}_0^1) \quad (48)$$

$$= P_{\underline{S}_1, Y_1 | \underline{S}_0}^{(\mathcal{P}_1)} (\underline{s}_1, y_1 | \underline{s}_0), \quad (49)$$

which directly implies

$$P_{\underline{S}_0^1, Y_1 | \underline{S}_0}^{(\mathcal{P}_2)} (\underline{s}_0^1, y_1 | \underline{s}_0) = P_{\underline{S}_0^1, Y_1 | \underline{S}_0}^{(\mathcal{P}_1)} (\underline{s}_0^1, y_1 | \underline{s}_0). \quad (50)$$

Now, assume that the equality (43) holds for up to time $t-1$, where $t > 1$, particularly,

$$P_{\underline{S}_{t-2}^{t-1}, Y_1^{t-1} | \underline{S}_0}^{(\mathcal{P}_2)} \left( \underline{s}_{t-2}^{t-1}, y_1^{t-1} | \underline{s}_0 \right)$$

$$= P_{\underline{S}_{t-2}^{t-1}, Y_1^{t-1} | \underline{S}_0}^{(\mathcal{P}_1)} \left( \underline{s}_{t-2}^{t-1}, y_1^{t-1} | \underline{s}_0 \right) \quad (51)$$

$$= \int \prod_{\tau=1}^{t-1} P_\tau \left( \underline{s}_\tau | \underline{s}_0^{\tau-1}, y_1^{\tau-1} \right) P_{Y_\tau | \underline{S}_{\tau-1}^\tau} \left( y_\tau | \underline{s}_{\tau-1}^\tau \right) \mathrm{d}\underline{s}_1^{t-3}. \quad (52)$$

The induction step for time $t$ is simply shown as follows

$$P_{\underline{S}_{t-1}^t, Y_1^t | \underline{S}_0}^{(\mathcal{P}_2)} \left( \underline{s}_{t-1}^t, y_1^t | \underline{s}_0 \right)$$

$$= Q_t(\underline{s}_t | \underline{s}_{t-1}, y_1^{t-1}) P_{Y_t | \underline{S}_{t-1}^t} (y_t | \underline{s}_{t-1}^t)$$

$$\int P_{\underline{S}_{t-2}^{t-1}, Y_1^{t-1} | \underline{S}_0}^{(\mathcal{P}_2)} \left( \underline{s}_{t-2}^{t-1}, y_1^{t-1} | \underline{s}_0 \right) \mathrm{d}\underline{s}_{t-2} \quad (53)$$

$$\stackrel{(a)}{=} \frac{\int \left[ \prod_{\tau=1}^{t-1} P_\tau(\underline{s}_\tau | \underline{s}_0^{\tau-1}, y_1^{\tau-1}) f_{Y_\tau | \underline{S}_{\tau-1}^\tau}(y_\tau | \underline{s}_{\tau-1}^\tau) \right] P_t(\underline{s}_t | \underline{s}_0^{t-1}, y_1^{t-1}) \mathrm{d}\underline{s}_1^{t-2}}{\int \left[ \prod_{\tau=1}^{t-1} P_\tau(\underline{s}_\tau | \underline{s}_0^{\tau-1}, y_1^{\tau-1}) f_{Y_\tau | \underline{S}_{\tau-1}^\tau}(y_\tau | \underline{s}_{\tau-1}^\tau) \right] \mathrm{d}\underline{s}_1^{t-2}} \times$$

$$P_{Y_t | \underline{S}_{t-1}^t}(y_t | \underline{s}_{t-1}^t) \int \left[ \prod_{\tau=1}^{t-1} P_\tau(\underline{s}_\tau | \underline{s}_0^{\tau-1}, y_1^{\tau-1}) f_{Y_\tau | \underline{S}_{\tau-1}^\tau}(y_\tau | \underline{s}_{\tau-1}^\tau) \right] \mathrm{d}\underline{s}_1^{t-2} \quad (54)$$

$$\stackrel{(b)}{=} \int \prod_{\tau=1}^t P_\tau \left( \underline{s}_\tau | \underline{s}_0^{\tau-1}, y_1^{\tau-1} \right) P_{Y_\tau | \underline{S}_{\tau-1}^\tau} \left( y_\tau | \underline{s}_{\tau-1}^\tau \right) \mathrm{d}\underline{s}_1^{t-2} \quad (55)$$

$$= P_{\underline{S}_{t-1}^t, Y_1^t | \underline{S}_0}^{(\mathcal{P}_1)} \left( \underline{s}_{t-1}^t, y_1^t | \underline{s}_0 \right), \quad (56)$$

where $(a)$ is the result of substituting the definition (39) for source $\mathcal{P}_2$ and the induction assumption (52) into (53), and $(b)$ is obtained by simplifying the expression in (54).

Thus, we have shown that the channel states $\underline{S}_{t-1}^t$ and outputs $Y_1^t$ induced by sources $\mathcal{P}_1$ and $\mathcal{P}_2$ have the same distribution. It is therefore clear that the sources $\mathcal{P}_1$ and $\mathcal{P}_2$ induce the same information $I(\mathcal{M}; Y_1^n | \underline{S}_0 = \underline{s}_0)$ according to (34).

Note that the set of channel state vector entries $\{\underline{S}_\tau(i) : 1 \leq i \leq M, 1 \leq \tau \leq t-1\}$ and the input sequence $X_1^{t-1}$ *linearly* determine each other according to the channel law in (19). Thus, induced by the Gaussian source $\mathcal{P}_1$, the channel state entries $\underline{S}_\tau(i)$ (for $1 \leq i \leq M$ and $1 \leq \tau \leq t$) and the symbols $Y_1^t$ are jointly Gaussian. We conclude that the conditional pdf's $Q_t(\underline{s}_t | \underline{s}_{t-1}, y_1^{t-1})$ constructed in (39) are also Gaussian functions and thus $\mathcal{P}_2$ is a feedback-dependent Gauss-Markov (not necessarily stationary) source. The Gauss-Markov source $\mathcal{P}_2$ also satisfies the input power constraint (22), which is obvious by the equality in (56). ∎

Theorem 1 reveals that, for any given prior channel output $y_1^{t-1}$, it is sufficient to utilize Markov sources to maximize the entropy of the channel output sequence. By Theorem 1, without loss of optimality, in the sequel we only consider feedback-dependent Gauss-Markov sources of the following form

$$\mathcal{P}^{\mathrm{GM}} \triangleq \left\{ P_t \left( \underline{s}_t | \underline{s}_{t-1}, y_1^{t-1} \right), t = 1, 2, \cdots \right\}. \quad (57)$$

### B. The Kalman-Bucy Filter is Optimal for Processing the Feedback

*Definition 1:* We use $\alpha_t(\cdot)$ as shorthand notation for the posterior pdf of the channel state $\underline{S}_t$ given the prior channel outputs $Y_1^t = y_1^t$, that is

$$\alpha_t(\underline{\mu}) \triangleq P_{\underline{S}_t | \underline{S}_0, Y_1^t} \left( \underline{\mu} | \underline{s}_0, y_1^t \right). \quad (58)$$

□

For a feedback-dependent Gauss-Markov source $\mathcal{P}^{\mathrm{GM}}$, the functions $\alpha_t(\cdot)$ can be recursively computed by the Bayes rule as follows

$$\alpha_t(\underline{\mu}) = \frac{\int \alpha_{t-1}(\underline{v}) P_t(\underline{\mu} | \underline{v}, y_1^{t-1}) P_{Y_t | \underline{S}_{t-1}, \underline{S}_t}(y_t | \underline{v}, \underline{\mu}) \mathrm{d}\underline{v}}{\iint \alpha_{t-1}(\underline{v}) P_t(\underline{u} | \underline{v}, y_1^{t-1}) P_{Y_t | \underline{S}_{t-1}, \underline{S}_t}(y_t | \underline{v}, \underline{u}) \mathrm{d}\underline{u} \mathrm{d}\underline{v}}. \quad (59)$$

Since the function $\alpha_t(\cdot)$ is a Gaussian pdf, it is completely characterized by the conditional mean $\underline{m}_t$ and conditional covariance matrix $\mathbf{K}_t$ of the channel state

$$\underline{m}_t = \mathrm{E}\left[\underline{S}_t | \underline{s}_0, y_1^t\right], \quad (60)$$

$$\mathbf{K}_t = \mathrm{E}\left[(\underline{S}_t - \underline{m}_t)(\underline{S}_t - \underline{m}_t)^{\mathrm{T}} | \underline{s}_0, y_1^t\right]. \quad (61)$$

We note that the recursion (59), i.e., the recursive computation of $\underline{m}_t$ and $\mathbf{K}_t$, can be implemented by a Kalman-Bucy filter [22].

*Theorem 2:* [*Kalman-Bucy filter is optimal for processing the feedback*] For the power-constrained linear Gaussian channel, let the feedback-dependent Gauss-Markov (not necessarily

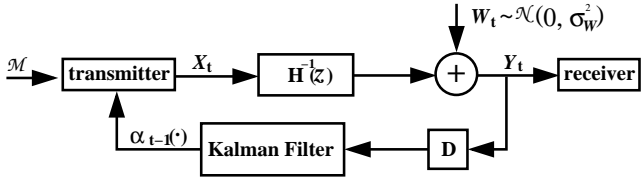

Fig. 7. The optimal feedback strategy for the Linear Gaussian noise channel.

stationary) source $\mathcal{P}_\alpha^{\text{GM}}$ be defined as

$$\mathcal{P}_\alpha^{\text{GM}} \triangleq \{P_t\left(\underline{s}_t \,|\, \underline{s}_{t-1}, \alpha_{t-1}(\cdot)\right), t = 1, 2, \ldots\}, \quad (62)$$

where the Markov transition probability depends only on the posterior distribution function of the channel state $\alpha_t(\cdot)$ instead of all prior channel outputs $Y_1^{t-1}$. The $n$-block feedback capacity $C^{\text{fb}(n)}$ then equals

$$C^{\text{fb}(n)} = \max_{\mathcal{P}_\alpha^{\text{GM}}} \frac{1}{n} I\left(\mathcal{M}; Y_1^n \,|\, \underline{S}_0 = \underline{s}_0\right)$$
$$= \max_{\mathcal{P}_\alpha^{\text{GM}}} \frac{1}{n} \sum_{t=1}^n I\left(\underline{S}_{t-1}^t; Y_t \,|\, Y_1^{t-1}, \underline{s}_0\right), \quad (63)$$

where the maximization is taken subject to an average input power constraint

$$\frac{1}{n} \sum_{t=1}^n \mathrm{E}\left[(X_t)^2 \,|\, \underline{S}_0 = \underline{s}_0\right] \leq P. \quad (64)$$

This capacity-achieving strategy is depicted in Figure 7. □

*Proof:* We first briefly outline the proof strategy. We will consider two *different* feedback vector realizations (channel output histories) $y_1^{t-1}$ and $\tilde{y}_1^{t-1}$ that both induce the same posterior state pdf $\alpha_{t-1}(\cdot)$ at time $t-1$. We will then apply equivalent Gauss-Markov source distributions for the subsequent source symbols (at times $t, t+1, \ldots$), irrespective of the feedback realization ($y_1^{t-1}$ or $\tilde{y}_1^{t-1}$), and we get the same distributions for the subsequent channel inputs and outputs, and thus the same transmission power and output entropy (for the subsequent transmissions), irrespective of the two channel histories. Using this result, we complete the proof using an inductive argument.

Suppose that two different feedback vectors $\tilde{y}_1^{t-1}$ and $y_1^{t-1}$ ($\tilde{y}_1^{t-1} \neq y_1^{t-1}$) induce the same posterior channel state pdf $\alpha_{t-1}(\cdot)$, that is, for any possible state value $\underline{s}_{t-1} = \underline{\mu}$ we have

$$\tilde{\alpha}_{t-1}\left(\underline{\mu}\right) \triangleq P_{\underline{S}_{t-1} | \underline{S}_0, Y_1^{t-1}}\left(\underline{\mu} \,|\, \underline{s}_0, \tilde{y}_1^{t-1}\right)$$
$$= P_{\underline{S}_{t-1} | \underline{S}_0, Y_1^{t-1}}\left(\underline{\mu} \,|\, \underline{s}_0, y_1^{t-1}\right) \triangleq \alpha_{t-1}\left(\underline{\mu}\right). \quad (65)$$

Now consider two distributions of the source $S_\tau$, for $\tau \geq t$, the first distribution conditioned on $y_1^{t-1}$, and the second conditioned on $\tilde{y}_1^{t-1}$. If we let these two distributions, irrespective of the feedback realization ($y_1^{t-1}$ or $\tilde{y}_1^{t-1}$), be equal to each other for $\tau \geq t$, that is, for $\tau \geq t$ if

$$P_\tau\left(\underline{s}_\tau \,|\, \underline{s}_{\tau-1}, \tilde{y}_1^{t-1}, y_t^{\tau-1}\right) = P_\tau\left(\underline{s}_\tau \,|\, \underline{s}_{\tau-1}, y_1^{t-1}, y_t^{\tau-1}\right), \quad (66)$$

we then have

$$P_{Y_t^n, \underline{S}_{t-1}^n | \underline{S}_0, Y_1^{t-1}}\left(y_t^n, \underline{s}_{t-1}^n \,|\, \underline{s}_0, \tilde{y}_1^{t-1}\right)$$
$$= \alpha_{t-1}(\underline{s}_{t-1}) \prod_{\tau=t}^n P_\tau(\underline{s}_\tau \,|\, \underline{s}_{\tau-1}, y_1^{\tau-1}) P_{Y_\tau | \underline{S}_{\tau-1}^\tau}(y_\tau \,|\, \underline{s}_{\tau-1}^\tau)$$
$$= P_{Y_t^n, \underline{S}_{t-1}^n | \underline{S}_0, Y_1^{t-1}}\left(y_t^n, \underline{s}_{t-1}^n \,|\, \underline{s}_0, y_1^{t-1}\right). \quad (67)$$

The equality in (67) directly implies that the entropies are equal

$$h\left(Y_t^n \,|\, \underline{s}_0, \tilde{y}_1^{t-1}\right) = h\left(Y_t^n \,|\, \underline{s}_0, y_1^{t-1}\right), \quad (68)$$

and that for any $\tau \geq t$ the transmission powers are equal

$$\mathrm{E}\left[(X_\tau)^2 \,|\, \underline{s}_0, \tilde{y}_1^{t-1}\right] = \mathrm{E}\left[(X_\tau)^2 \,|\, \underline{s}_0, y_1^{t-1}\right]. \quad (69)$$

Therefore, for any $t > 0$ and any constant $\Pi \geq 0$, the source distribution $P_\tau\left(\underline{s}_\tau \,|\, \underline{s}_{\tau-1}, y_1^{t-1}, y_t^{\tau-1}\right)$ for time $t \leq \tau \leq n$ that maximizes the channel output entropy $h\left(Y_t^n \,|\, \underline{s}_0, y_1^{t-1}\right)$, subject to power constraint

$$\sum_{\tau=t}^n \mathrm{E}\left[(X_\tau)^2 \,|\, \underline{s}_0, y_1^{t-1}\right] \leq \Pi, \quad (70)$$

when $y_1^{t-1}$ is the feedback vector, must also maximize the entropy $h\left(Y_t^n \,|\, \underline{s}_0, \tilde{y}_1^{t-1}\right)$ under constraint

$$\sum_{\tau=t}^n \mathrm{E}\left[(X_\tau)^2 \,|\, \underline{s}_0, \tilde{y}_1^{t-1}\right] \leq \Pi, \quad (71)$$

when $\tilde{y}_1^{t-1}$ is the feedback vector. We note that $\Pi$ is an arbitrary non-negative constant (independent of the channel output history, and independent of the initial state).

According to the above analysis, for any given power budget $\Pi \geq 0$ for transmissions at times $\tau \geq t$, the optimal source distribution $P_t\left(\underline{s}_t \,|\, \underline{s}_{t-1}, y_1^{t-1}\right)$ at time $t$ that maximizes the entropy of subsequent channel outputs depends only on $\alpha_{t-1}(\cdot)$. We may summarize the conclusion as follows:

> **Conclusion 1:** At any time instant $t$, for any constant non-negative power constraint $\Pi$, the optimal distribution of the source $X_t^n$ that maximizes the channel output entropy $h\left(Y_t^n \,|\, \underline{s}_0, y_1^{t-1}\right)$ under the power constraint $\sum_{\tau=t}^n \mathrm{E}\left[(X_\tau)^2 \,|\, \underline{s}_0, y_1^{t-1}\right] \leq \Pi$, depends only on the posterior channel state distribution $\alpha_{t-1}(\cdot)$, regardless of the channel output history $y_1^{t-1}$ that lead to that posterior state distribution.

We now show that we can drop the conditioning on the channel output history $y_1^{t-1}$ when formulating the power constraint in Conclusion 1. According to (29), the maximization problem in (63) is equivalent to maximizing the channel output entropy, i.e.,

$$\arg\max_{\mathcal{P}} \sum_{t=1}^n I\left(\underline{S}_{t-1}^t; Y_t \,|\, Y_1^{t-1}, \underline{s}_0\right)$$
$$= \arg\max_{\mathcal{P}} h\left(Y_1^n \,|\, \underline{s}_0\right). \quad (72)$$

For any $1 \leq t \leq n$, we can rewrite the channel output entropy as

$$h\left(Y_1^n \,|\, \underline{s}_0\right) = h\left(Y_1^{t-1} \,|\, \underline{s}_0\right) +$$
$$\int h\left(Y_t^n \,|\, \underline{s}_0, y_1^{t-1}\right) P_{Y_1^{t-1}}\left(y_1^{t-1}\right) \mathrm{d}y_1^{t-1}. \quad (73)$$

Similarly, we can rewrite the input power as

$$\sum_{\tau=1}^{n} \mathrm{E}\left[(X_\tau)^2 \,|\, \underline{s}_0\right] = \sum_{\tau=1}^{t-1} \mathrm{E}\left[(X_\tau)^2 \,|\, \underline{s}_0\right] + \int \sum_{\tau=t}^{n} \mathrm{E}\left[(X_\tau)^2 \,|\, \underline{s}_0, y_1^{t-1}\right] P_{Y_1^{t-1}}\left(y_1^{t-1}\right) \mathrm{d} y_1^{t-1}. \quad (74)$$

For the optimal source distribution that maximizes (72) subject to the power constraint (64), and for any given channel output realization $y_1^{t-1}$, we define the remaining power at time $t$ as

$$\Pi_{t-1}\left(y_1^{t-1}\right) \triangleq \sum_{\tau=t}^{n} \mathrm{E}\left[(X_\tau)^2 \,|\, \underline{s}_0, y_1^{t-1}\right]. \quad (75)$$

Clearly $\Pi_0 = nP$. By definition, $\Pi_{t-1}\left(y_1^{t-1}\right)$ is a function of prior channel outputs and a function of time $t$. We note that the channel inputs and outputs are jointly Gaussian [7]. Thus for any $\tau \geq t$, it is sufficient to only consider optimal Gaussian sources that satisfy

$$\mathrm{E}\left[X_\tau \,|\, \underline{s}_0, y_1^{t-1}\right] = 0, \quad (76)$$

because otherwise one could easily verify that another Gaussian source defined as $\hat{X}_\tau \triangleq X_\tau - \mathrm{E}\left[X_\tau \,|\, \underline{s}_0, y_1^{t-1}\right]$ for $\tau \geq t$ would induce the same channel output entropy (73) while consuming strictly less power than $X_\tau$. Now, since $\mathrm{E}\left[X_\tau \,|\, \underline{s}_0, y_1^{t-1}\right] = 0$, we have

$$\sum_{\tau=t}^{n} \mathrm{E}\left[(X_\tau)^2 \,|\, \underline{s}_0, y_1^{t-1}\right] = \sum_{\tau=t}^{n} \mathrm{Var}\left(X_\tau \,|\, \underline{s}_0, y_1^{t-1}\right). \quad (77)$$

But since $X_\tau$ and $Y_1^{t-1}$ are jointly Gaussian, the variance $\mathrm{Var}\left(X_\tau \,|\, \underline{s}_0, y_1^{t-1}\right)$ does not depend on the realization $y_1^{t-1}$, and we can write

$$\sum_{\tau=t}^{n} \mathrm{E}\left[(X_\tau)^2 \,|\, \underline{s}_0, y_1^{t-1}\right] = \sum_{\tau=t}^{n} \mathrm{E}\left[(X_\tau)^2 \,|\, \underline{s}_0\right], \quad (78)$$

or equivalently, $\Pi_{t-1} = \Pi_{t-1}\left(y_1^{t-1}\right)$. Hence, conditioning on the channel output history $y_1^{t-1}$, is irrelevant when formulating the power constraint. Directly utilizing (78), we can modify Conclusion 1 into a new (slightly relaxed) conclusion that does not require conditioning on the channel output history when formulating the power constraint.

**Conclusion 2:** At any time instant $t$, for any constant non-negative power constraint $\Pi$, the optimal distribution of the source $X_t^n$ that maximizes the channel output entropy $h\left(Y_t^n \,|\, \underline{s}_0, y_1^{t-1}\right)$ under the power constraint $\sum_{\tau=t}^{n} \mathrm{E}\left[(X_\tau)^2 \,|\, \underline{s}_0, y_1^{t-1}\right] = \sum_{\tau=t}^{n} \mathrm{E}\left[(X_\tau)^2 \,|\, \underline{s}_0\right] \leq \Pi$, depends only on the posterior channel state distribution $\alpha_{t-1}(\cdot)$, regardless of the channel output history $y_1^{t-1}$ that lead to that posterior state distribution.

We now utilize Conclusion 2 to finalize the proof using an inductive argument. Let $\Pi_0 = nP$ be the power constraint at the beginning of transmissions. We have already established in Conclusion 2 that given the power constraint $\Pi_0$, the optimal source that maximizes $h\left(Y_1^n \,|\, \underline{s}_0\right)$ under the power constraint $\Pi_0$ depends only on $\alpha_0(\cdot)$.[2] The optimal source now generates the channel input $X_1$ at time $t = 1$, whose power is $\pi_1 = \mathrm{E}\left[(X_1)^2 \,|\, \underline{s}_0\right]$. The channel input $X_1$ induces the channel output $Y_1$, which in turn produces a new posterior state distribution $\alpha_1(\cdot)$. The leftover power is $\Pi_1 = \Pi_0 - \pi_1$. Now, the optimal source at time instant $t = 2$ is one that maximizes $h\left(Y_2^n \,|\, \underline{s}_0, y_1\right)$ subject to the power constraint $\Pi_1$ and, according to Conclusion 2, the source distribution at time $t = 2$ depends only on $\alpha_1(\cdot)$. This source generates $X_2$. Extending the inductive argument further in similar fashion, we conclude that at each time step $t$, according to Conclusion 2, the source distribution at time $t$ depends only on the posterior state distribution $\alpha_{t-1}(\cdot)$. ∎

An alternative proof based on dynamic programming [23] and Lagrange multipliers is given in Appendix 1.

Theorem 2 suggests that, for the task of constructing the next signal to be transmitted, all the "knowledge" contained in the vector of prior channel outputs $y_1^{t-1}$ is captured by the posterior pdf of the channel state $\alpha_{t-1}(\cdot)$.

So far, we have simplified the capacity-achieving feedback strategy such that the transmitter does not need to memorize the entire channel dynamics $y_1^{t-1}$. Instead, for forming the optimal signal $X_t$ (or equivalently state $\underline{S}_t$) to be transmitted at time $t$, we only need to know the immediately preceding channel state $\underline{S}_{t-1} = \underline{s}_{t-1}$ (determined by the prior channel inputs) and the Kalman-Bucy filter output, which is a Gaussian probability density function $\alpha_{t-1}(\cdot)$ characterized by the conditional mean $\underline{m}_{t-1}$ and the conditional covariance matrix $\mathbf{K}_{t-1}$.

### C. Properties of Capacity-Achieving Channel Dynamics

For any feedback-dependent Gauss-Markov source $\mathcal{P}_\alpha^{\mathrm{GM}}$, we have the following properties for the channel states, outputs and posterior state distributions.

*Theorem 3:* For the linear Gaussian noise channel, if the feedback-dependent Gauss-Markov source distribution $\mathcal{P}_\alpha^{\mathrm{GM}}$ is used, we have

1) The sequence of posterior channel state distributions $\alpha_t(\cdot)$, or the pairs of variables $\underline{m}_t$ and $\mathbf{K}_t$, is a Markov random process.
2) The sequence of the pairs $(\alpha_t(\cdot), \underline{S}_t)$ is a Markov random process.
3) The sequence of the pairs $(\alpha_t(\cdot), Y_t)$ is a Markov random process.
4) The sequence of the triples $(\alpha_t(\cdot), \underline{S}_t, Y_t)$ is a Markov random process. □

*Proof:* For the Gauss-Markov source $\mathcal{P}_\alpha^{\mathrm{GM}}$, the recursive sum-product rule (59) for updating the posterior pdf of the channel state becomes

$$\alpha_t(\underline{\mu}) = \frac{\int \alpha_{t-1}(\underline{v}) P_t(\underline{\mu} | \underline{v}, \alpha_{t-1}(\cdot)) P_{Y_t | \underline{S}_{t-1}, \underline{S}_t}(y_t | \underline{v}, \underline{\mu}) \mathrm{d}\underline{v}}{\iint \alpha_{t-1}(\underline{v}) P_t(\underline{u} | \underline{v}, \alpha_{t-1}(\cdot)) P_{Y_t | \underline{S}_{t-1}, \underline{S}_t}(y_t | \underline{v}, \underline{u}) \mathrm{d}\underline{u}\mathrm{d}\underline{v}} \quad (79)$$

---

[2]Note that here $\alpha_0(\underline{\mu}) = \delta(\underline{\mu} - \underline{s}_0)$ because $\underline{s}_0$ is the known initial state.

and the conditional pdf of channel output $y_t$ is

$$P_{Y_t|\alpha_{t-1}(\cdot)}\left(y_t \,|\alpha_{t-1}(\cdot)\right) = P_{Y_t|\underline{S}_0,Y_1^{t-1}}\left(y_t \,|\underline{s}_0, y_1^{t-1}\right)$$
$$= \iint \alpha_{t-1}(\underline{v}) P_t(\underline{u}|\underline{v},\alpha_{t-1}(\cdot)) P_{Y_t|\underline{S}_{t-1},\underline{S}_t}(y_t|\underline{v},\underline{u}) \,\mathrm{d}\underline{u}\mathrm{d}\underline{v}. \quad (80)$$

We note that in (79) and (80), the dependence on $\underline{s}_0$ and $y_1^{t-1}$ is replaced by the dependence on $\alpha_{t-1}(\cdot)$ according to Theorem 2.

The theorem follows from (79), (80) and the Markovianity of the source $\mathcal{P}_\alpha^{\mathrm{GM}}$. ■

### D. Feedback-Capacity-Achieving Sources for General State-Machine Channels

Theorems 1, 2 and 3 also hold for more general state-machine (or state-space) channels other than the linear Gaussian noise channel, except that for general state-machine channels, the feedback-capacity-achieving source may not be Gaussian any more, and that the Kalman-Bucy filter is replaced by a discrete-time version of the Wonham filter [24] (or the forward sum-product recursion of the BCJR or Baum-Welch algorithm [25], [26], [27], [28] for finite-state machines, see [29]).

The linear Gaussian noise channel in this paper has an average input power constraint over the whole block (or horizon), which makes the feedback capacity computation problem even more difficult than the state-machine channel considered in [29], where no average power constraint is needed and the inputs are chosen from a finite-size alphabet. We note that the arguments used to prove Theorems 1, 2 and 3 also hold for some other channel input power constraints, e.g., the peak input power constraint, which is beyond the scope of this paper and will not be elaborated on any further.

## V. $n$-Block Feedback Capacity Computation

### A. Parameterizing the Feedback-Capacity-Achieving Markov Sources

*Lemma 1:* Without loss of generality, the Gauss-Markov source $\mathcal{P}_\alpha^{\mathrm{GM}}$ can be expressed as

$$X_t = \underline{d}_t^{\mathrm{T}} \underline{S}_{t-1} + e_t Z_t + g_t, \quad (81)$$

where $Z_t$ is a white Gaussian random process with unit-variance and is independent of $X_1^{t-1}$ and $Y_1^{t-1}$. (Here $e_t$ and $g_t$ are scalars, and $\underline{d}_t$ is a vector of length $L$.) The coefficients $\underline{d}_t$, $e_t$ and $g_t$ are all dependent on the Gaussian pdf $\alpha_{t-1}(\cdot)$, or alternatively on its mean $\underline{m}_{t-1}$ and covariance matrix $\mathbf{K}_{t-1}$. The set of coefficients $\{\underline{d}_t, e_t, g_t\}$ completely determines the feedback-dependent Gauss-Markov source distribution $\mathcal{P}_\alpha^{\mathrm{GM}}$ needed in Theorem 2. □

*Proof:* Given the channel state realization $\underline{S}_{t-1} = \underline{s}_{t-1}$, the input $x_t$ and state $\underline{s}_t$ determine each other uniquely by the channel state propagation rule (19), so the feedback-dependent Gauss-Markov source $\mathcal{P}_\alpha^{\mathrm{GM}}$ in Theorem 2 can be equivalently represented as

$$\mathcal{P}_\alpha^{\mathrm{GM}} = \left\{ P_t\left(x_t \,|\underline{s}_{t-1}, \alpha_{t-1}(\cdot)\right), t = 1, 2, \ldots \right\}. \quad (82)$$

Further, since the channel state propagation rule (19) is linear, the joint distribution of $X_t$ and $\underline{S}_{t-1}$ is also Gaussian for any given $\alpha_{t-1}(\cdot)$. Thus, without loss of generality, the source $\mathcal{P}_\alpha^{\mathrm{GM}}$ can be specified as in (81) by coefficients $\underline{d}_t, e_t$ and $g_t$, which depend on the Kalman-Bucy filter output $\alpha_{t-1}(\cdot)$. ■

The source parametrization in (81) reveals further structure when compared to the source parametrization (14) obtained by Cover and Pombra [7]. It is clear that every parametrization (81) leads to an equivalent parametrization (14), but not vice versa. Also note that the number of parameters $\underline{d}_t$, $e_t$ and $g_t$ in (81) grows only linearly with the horizon distance $n$, while the number of parameters $b_{it}$ in (14) grows quadratically with the horizon distance $n$. The encoding complexity in (14) grows linearly with time in general, while the encoding complexity in (81) is constant for any time instant $t > 1$.

The following lemma establishes a formula for each term in the information sum (63) and the input signal power sum (64) in terms of the source parameters $\underline{d}_t$, $e_t$, and $g_t$.

*Lemma 2:* For the feedback-dependent Gauss-Markov source (81), we have

$$h\left(Y_t \,|\underline{s}_0, y_1^{t-1}\right) - \frac{1}{2}\log\left(2\pi e \sigma_W^2\right)$$
$$= \frac{1}{2}\log\left(\frac{\sigma_W^2 + (\underline{a}+\underline{c}+\underline{d}_t)^{\mathrm{T}}\mathbf{K}_{t-1}(\underline{a}+\underline{c}+\underline{d}_t) + (e_t)^2}{\sigma_W^2}\right), \quad (83)$$

and

$$\mathrm{E}\left[(X_t)^2 \,|\underline{s}_0, y_1^{t-1}\right] = \left(\underline{d}_t^{\mathrm{T}} \underline{m}_{t-1} + g_t\right)^2 + \underline{d}_t^{\mathrm{T}} \mathbf{K}_{t-1} \underline{d}_t + (e_t)^2, \quad (84)$$

where $\underline{d}_t$, $e_t$ and $g_t$ are themselves functions of $\underline{m}_{t-1}$ and $\mathbf{K}_{t-1}$. □

*Proof:* For the source (81) in Lemma 1, using the Bayes rule, we can formulate the first and second order conditional moments of the channel input $X_t$ and channel output $Y_t$ in terms of $\underline{d}_t$, $\underline{e}_t$, $g_t$, $\underline{m}_{t-1}$ and $\mathbf{K}_{t-1}$ as

$$\mathrm{E}\left[X_t \,|\underline{s}_0, y_1^{t-1}\right] = \underline{d}_t^{\mathrm{T}} \underline{m}_{t-1} + g_t, \quad (85)$$
$$\mathrm{E}\left[(X_t)^2 \,|\underline{s}_0, y_1^{t-1}\right]$$
$$= \left(\underline{d}_t^{\mathrm{T}} \underline{m}_{t-1} + g_t\right)^2 + \underline{d}_t^{\mathrm{T}} \mathbf{K}_{t-1} \underline{d}_t + (e_t)^2, \quad (86)$$
$$\mathrm{E}\left[Y_t \,|\underline{s}_0, y_1^{t-1}\right] = (\underline{a}+\underline{c}+\underline{d}_t)^{\mathrm{T}} \underline{m}_{t-1} + g_t, \quad (87)$$
$$\mathrm{E}\left[\left(Y_t - \mathrm{E}\left[Y_t \,|\underline{s}_0, y_1^{t-1}\right]\right)^2 \,|\underline{s}_0, y_1^{t-1}\right]$$
$$= (\underline{a}+\underline{c}+\underline{d}_t)^{\mathrm{T}} \mathbf{K}_{t-1} (\underline{a}+\underline{c}+\underline{d}_t) + (e_t)^2 + \sigma_W^2. \quad (88)$$

Note that conditioned on $\underline{s}_0$ and $y_1^{t-1}$, the random variable $Y_t$ has a Gaussian distribution with variance as in (88), thus we obtain (83). Equation (84) is the same as (86). ■

### B. Problem Reformulation Using Lagrange Multipliers

The feedback capacity computation problem as stated in Theorem 2 is a maximization problem under an inequality constraint. Since the $n$-block feedback capacity computation problem is a convex optimization problem [19], and since Slater's conditions (see [30], Section 5.2.3) for this optimization problem are satisfied, strong duality holds between the original optimization problem and its Lagrangian dual. Such optimization problems can be reformulated (or solved) by using the method of Lagrange multipliers.

We first define a reward function for this optimization problem.

*Definition 2:* For an arbitrary constant $\gamma \geq 0$, we define the reward function $\Omega(\cdot)$ as

$$\Omega\left(\underline{m}_{t-1}, \mathbf{K}_{t-1}, \underline{d}_t, e_t, g_t, \gamma\right)$$
$$\triangleq \underbrace{h\left(Y_t \mid \underline{s}_0, y_1^{t-1}\right) - \frac{1}{2}\log\left(2\pi e \sigma_W^2\right)}_{\text{information transmitted}} - \underbrace{\gamma \mathrm{E}\left[(X_t)^2 \mid \underline{s}_0, y_1^{t-1}\right]}_{\text{penalty for using power}}$$

$$\stackrel{(a)}{=} \frac{1}{2}\log\left(\frac{\sigma_W^2 + (\underline{a}+\underline{c}+\underline{d}_t)^{\mathrm{T}}\mathbf{K}_{t-1}(\underline{a}+\underline{c}+\underline{d}_t)+(e_t)^2}{\sigma_W^2}\right)$$
$$- \gamma\left(\left(\underline{d}_t^{\mathrm{T}}\underline{m}_{t-1} + g_t\right)^2 + \underline{d}_t^{\mathrm{T}}\mathbf{K}_{t-1}\underline{d}_t + (e_t)^2\right), \quad (89)$$

where equality $(a)$ follows from Lemma 2. □

*Theorem 4:* For a linear Gaussian noise channel with noiseless feedback, the $n$-block feedback capacity under the average input power constraint equals

$$C^{\mathrm{fb}(n)} = C^{\mathrm{fb}(n)}(P)$$
$$= \max_{\mathcal{P}_\alpha^{\mathrm{GM}}} \frac{1}{n}\mathrm{E}\left[\sum_{t=1}^n \Omega\left(\underline{m}_{t-1}, \mathbf{K}_{t-1}, \underline{d}_t, e_t, g_t, \gamma\right)\right] + \gamma P, \quad (90)$$

where $\gamma \geq 0$ and

$$\gamma\left(\frac{1}{n}\sum_{t=1}^n \mathrm{E}\left[(X_t)^2 \mid \underline{S}_0 = \underline{s}_0\right] - P\right) = 0 \quad (91)$$

are first-order Kuhn-Tucker necessary conditions for achieving the feedback capacity. □

*Proof:* It has already been established that the $n$-block feedback capacity computation is a convex optimization problem [19]. It is easy to verify that Slater's conditions (see [30], Section 5.2.3) hold, hence the primal and the Lagrangian dual problems are equivalent (see Appendix 2-A and Appendix 2-C for a complete proof). Thus, we can focus on the dual Lagrangian problem to derive the optimal source distribution so as to achieve and compute the $n$-block feedback capacity.

We consider the optimization problem in Theorem 2, i.e., we consider finding

$$\max_{\mathcal{P}_\alpha^{\mathrm{GM}}} \sum_{t=1}^n \mathrm{E}_{Y_1^{t-1}}\left[h\left(Y_t \mid \underline{s}_0, y_1^{t-1}\right) - \frac{1}{2}\log\left(2\pi e \sigma_W^2\right)\right], \quad (92)$$

subject to

$$\sum_{t=1}^n \mathrm{E}\left[(X_t)^2 \mid \underline{S}_0 = \underline{s}_0\right] \leq nP. \quad (93)$$

The Lagrangian (with Lagrange multiplier $\gamma$) is

$$\mathcal{L}^{(n)}\left(\mathcal{P}_\alpha^{\mathrm{GM}}, \gamma\right)$$
$$= \sum_{t=1}^n \mathrm{E}_{Y_1^{t-1}}\left[h\left(Y_t \mid \underline{s}_0, y_1^{t-1}\right) - \frac{1}{2}\log\left(2\pi e \sigma_W^2\right)\right]$$
$$- \gamma\left(\sum_{t=1}^n \mathrm{E}\left[(X_t)^2 \mid \underline{S}_0 = \underline{s}_0\right] - nP\right) \quad (94)$$

$$= \mathrm{E}_{Y_1^{n-1}}\left[\sum_{t=1}^n \left(h\left(Y_t \mid \underline{s}_0, y_1^{t-1}\right) - \frac{1}{2}\log\left(2\pi e \sigma_W^2\right)\right.\right.$$
$$\left.\left. - \gamma \mathrm{E}\left[(X_t)^2 \mid \underline{s}_0, y_1^{t-1}\right]\right)\right] + n\gamma P \quad (95)$$

$$= \mathrm{E}\left[\sum_{t=1}^n \Omega\left(\underline{m}_{t-1}, \mathbf{K}_{t-1}, \underline{d}_t, e_t, g_t, \gamma\right)\right] + n\gamma P. \quad (96)$$

The source $\mathcal{P}_\alpha^{\mathrm{GM}}$ is optimal for the horizon $n$ if and only if it maximizes the Lagrangian $\mathcal{L}^{(n)}\left(\mathcal{P}_\alpha^{\mathrm{GM}}, \gamma\right)$ and also satisfies the first-order Kuhn-Tucker necessary conditions [31], which are $\gamma \geq 0$ and

$$\gamma\left(\sum_{t=1}^n \mathrm{E}\left[(X_t)^2 \mid \underline{S}_0 = \underline{s}_0\right] - nP\right) = 0. \quad (97)$$

Further, when the maximum in (92) is achieved, since equality (97) holds, we have

$$\max_{\mathcal{P}_\alpha^{\mathrm{GM}}} \mathcal{L}^{(n)}\left(\mathcal{P}_\alpha^{\mathrm{GM}}, \gamma\right)$$
$$\stackrel{(a)}{=} \max_{\mathcal{P}_\alpha^{\mathrm{GM}}} \sum_{t=1}^n \mathrm{E}_{Y_1^{t-1}}\left[h\left(Y_t \mid \underline{s}_0, y_1^{t-1}\right) - \frac{1}{2}\log\left(2\pi e \sigma_W^2\right)\right] \quad (98)$$
$$= \max_{\mathcal{P}_\alpha^{\mathrm{GM}}} I\left(\mathcal{M}; Y_1^n \mid \underline{S}_0 = \underline{s}_0\right), \quad (99)$$

where $(a)$ is obtained by substituting (97) into (94). ∎

Occasionally, the Lagrange multiplier $\gamma$ is called as the *shadow price* [32] for the optimization problem. As can be seen in Definition 2, the value of $\gamma$ determines the penalty that incurs for using the signal power in the objective function (90). It can be verified that the shadow price $\gamma$ and the power constraint $P$ are in a 1-to-1 correspondence (see Corollary F in Appendix 2-D), and that $\gamma$ monotonically decreases with $P$ (see Proposition E in Appendix 2-D). Particularly, the value $\gamma = 0$ corresponds to the case when there is no power constraint, or $P = \infty$. In this paper, we are interested in a finite input power constraint, i.e., $P < \infty$, thus in the sequel we only consider $\gamma > 0$.

*Corollary 4.1:* For any finite power budget $P$, the corresponding Lagrange multiplier (or shadow price) satisfies $\gamma > 0$. The feedback-dependent Gauss-Markov source $\mathcal{P}^*$ that satisfies

$$\mathcal{P}^* = \arg\max_{\mathcal{P}_\alpha^{\mathrm{GM}}} \frac{1}{n}\mathrm{E}\left[\sum_{t=1}^n \Omega\left(\underline{m}_{t-1}, \mathbf{K}_{t-1}, \underline{d}_t, e_t, g_t, \gamma\right)\right] \quad (100)$$

achieves the $n$-block feedback capacity $C^{\mathrm{fb}(n)}$, where the power budget $P$ satisfies the following equality

$$P = \frac{1}{n}\sum_{t=1}^n \mathrm{E}\left[(X_t)^2 \mid \underline{S}_0 = \underline{s}_0\right]. \quad (101)$$

Here, the condition in (101) is the optimal power configuration. □

*Proof:* For $\gamma > 0$, the condition (91) in Theorem 4 becomes (101). Thus by Theorem 4 (see also Appendix 2-D) the source $\mathcal{P}^*$ that achieves the $n$-block feedback capacity $C^{\mathrm{fb}(n)} = C^{\mathrm{fb}(n)}(P)$ in (90) has power equal to $P$ and satisfies (101). ∎

Corollary 4.1 asserts that we can substitute the inequality power constraint (22) by the corresponding *equality*, and for any horizon $n$ the power budget $P > 0$ corresponds to a shadow price $\gamma$. For any shadow price $\gamma > 0$, the source determined in Corollary 4.1 is optimal for the specific power budget $P$ satisfying (101); see Appendix 2-D for the proof. In other words, the power budget $P$ is a monotonic (see Proposition E in Appendix 2-D), though not explicitly expressible, function of $\gamma$ (see also [33], [32]). Thus, computing the $n$-block feedback capacity $C^{\mathrm{fb}(n)}$ for a finite power budget $0 < P < \infty$ is equivalent to solving the maximization problem posed in Corollary 4.1 for some positive shadow price $\gamma > 0$ (as proved in Appendix 2-D), although the relationship between $P$ and $\gamma$ is not known in an explicit form.

We note that the general relationship between the power budget $P$ and the shadow price $\gamma$ may also be derived from [7], but the explicitly separable form of the objective function in Theorem 4 and Corollary 4.1 is a new result and will be crucial for obtaining more explicit solutions to the $n$-block feedback capacity computation problem. It should also be mentioned that it was shown in [33] that the feedback capacity is concave in the power budget $P$.

We next study and solve the maximization problem in Corollary 4.1 for any given value of the shadow price $\gamma > 0$.

### C. Optimal Stochastic Control Formulation

The problem of finding the optimal source for a given value of $\gamma$ (see Corollary 4.1), can be formulated as a standard dynamic-system stochastic control problem [23] (Vol 1, Chapter 7 and Vol 2). We describe the dynamic system as follows.

The *state* of the dynamic system at each stage (or time) $t$ is the posterior pdf $\alpha_{t-1}(\cdot)$, which is characterized by its mean $\underline{m}_{t-1}$ and covariance matrix $\mathbf{K}_{t-1}$. The *control* or (*policy*) for stage $t$ is the Gauss-Markov (not necessarily stationary) source distribution function $P_t\left(\underline{s}_t|\underline{s}_{t-1}, \alpha_{t-1}(\cdot)\right)$ characterized by the parameters $\underline{d}_t$, $e_t$ and $g_t$ (see Lemma 1), which themselves are functions of $\underline{m}_{t-1}$ and $\mathbf{K}_{t-1}$. The system *disturbance* at stage $t$ is the noisy channel output $Y_t$, which has the following distribution

$$P_{Y_t|\alpha_{t-1}(\cdot)}\left(y_t|\alpha_{t-1}(\cdot)\right) = P_{Y_t|\underline{S}_0, Y_1^{t-1}}\left(y_t|\underline{s}_0, y_1^{t-1}\right)$$
$$= \iint \alpha_{t-1}(\underline{v}) P_t(\underline{u}|\underline{v}, \alpha_{t-1}(\cdot)) P_{Y_t|\underline{S}_{t-1}, \underline{S}_t}(y_t|\underline{v}, \underline{u})\, d\underline{u}\, d\underline{v}. \quad (102)$$

For this dynamic system which has $\alpha_{t-1}(\cdot)$ as the state, $P_t\left(\underline{s}_t|\underline{s}_{t-1}, \alpha_{t-1}(\cdot)\right)$ as the control and $Y_t$ as the disturbance, the system equation is

$$\alpha_t(\underline{\mu}) = \frac{\int \alpha_{t-1}(\underline{v}) P_t(\underline{\mu}|\underline{v}, \alpha_{t-1}(\cdot)) P_{Y_t|\underline{S}_{t-1}, \underline{S}_t}(y_t|\underline{v}, \underline{\mu})\, d\underline{v}}{\iint \alpha_{t-1}(\underline{v}) P_t(\underline{u}|\underline{v}, \alpha_{t-1}(\cdot)) P_{Y_t|\underline{S}_{t-1}, \underline{S}_t}(y_t|\underline{v}, \underline{u})\, d\underline{u}\, d\underline{v}}, \quad (103)$$

which can be implemented by the Kalman-Bucy filter, symbolically expressed as

$$(\underline{m}_t, \mathbf{K}_t) = F_{KB}^{(\underline{m}, \mathbf{K})}\left(\underline{m}_{t-1}, \mathbf{K}_{t-1}, \underline{d}_t, e_t, g_t, y_t\right). \quad (104)$$

If we define the matrix $\mathbf{Q}_t$ using $\mathbf{A}$ and $\underline{b}$ in (21) as

$$\mathbf{Q}_t \triangleq \mathbf{A} + \underline{b}\, \underline{d}_t^{\mathrm{T}}, \quad (105)$$

then, we can write the component-wise Kalman-Bucy filter (104) more explicitly as a pair of propagation equations [22]

$$\underline{m}_t = F_{KB}^{(\underline{m})}\left(\underline{m}_{t-1}, \mathbf{K}_{t-1}, \underline{d}_t, e_t, g_t, y_t\right)$$
$$= \mathbf{Q}_t \underline{m}_{t-1} + g_t \underline{b} +$$
$$\frac{\left(\mathbf{Q}_t \mathbf{K}_{t-1}(\underline{a}+\underline{c}+\underline{d}_t) + \underline{b}(e_t)^2\right)\left(y_t - (\underline{a}+\underline{d}_t)^{\mathrm{T}} \underline{m}_{t-1}\right)}{(\underline{a}+\underline{c}+\underline{d}_t)^{\mathrm{T}} \mathbf{K}_{t-1}(\underline{a}+\underline{c}+\underline{d}_t) + (e_t)^2 + \sigma_W^2}, \quad (106)$$

and

$$\mathbf{K}_t = F_{KB}^{(\mathbf{K})}\left(\mathbf{K}_{t-1}, \underline{d}_t, e_t\right)$$
$$= \mathbf{Q}_t \mathbf{K}_{t-1} \mathbf{Q}_t^{\mathrm{T}} + \underline{b}\,\underline{b}^{\mathrm{T}}(e_t)^2$$
$$- \frac{\left(\mathbf{Q}_t \mathbf{K}_{t-1}(\underline{a}+\underline{c}+\underline{d}_t) + \underline{b}(e_t)^2\right)\left(\mathbf{Q}_t \mathbf{K}_{t-1}(\underline{a}+\underline{c}+\underline{d}_t) + \underline{b}(e_t)^2\right)^{\mathrm{T}}}{(\underline{a}+\underline{c}+\underline{d}_t)^{\mathrm{T}} \mathbf{K}_{t-1}(\underline{a}+\underline{c}+\underline{d}_t) + (e_t)^2 + \sigma_W^2}. \quad (107)$$

Here, we note that $\mathbf{K}_t$ is a deterministic function of $\mathbf{K}_{t-1}$, $\underline{d}_t$ and $e_t$, and can be computed offline.

By Theorem 3, when the feedback-dependent Gauss-Markov source $\mathcal{P}_\alpha^{\mathrm{GM}}$ is used, the process $\alpha_{t-1}(\cdot)$ has a Markov property. For this dynamic system, if we define the *reward* for each stage $t$ as $\Omega\left(\underline{m}_{t-1}, \mathbf{K}_{t-1}, \underline{d}_t, e_t, g_t, \gamma\right)$, then the source optimization problem in Corollary 4.1 is an *average-reward-per-stage* stochastic control problem (see the *average-cost-per-stage* stochastic control problem in [23], Chapter 7 and Volume 2).

### D. Source Optimization and Feedback Capacity Computation

We now describe a fairly simple dynamic-programming value-iteration algorithm [23] which finds, for any horizon $n$, the optimal source distribution that maximizes the information $I\left(\mathcal{M}; Y_1^n | \underline{S}_0 = \underline{s}_0\right)$, and thus computes the $n$-block feedback capacity $C^{\mathrm{fb}(n)}$ for any shadow price $\gamma$ (or equivalently for the corresponding power budget $P$).

**Algorithm 1** FOR OPTIMIZING THE FEEDBACK-CAPACITY-ACHIEVING SOURCE DISTRIBUTION FOR ANY GIVEN SHADOW PRICE $\gamma > 0$

---

**Initialization:** For any possible value of the pair $(\underline{m}_{t-1}, \mathbf{K}_{t-1})$, define the *terminal reward function* as $J^{(0)}\left(\underline{m}_{t-1}, \mathbf{K}_{t-1}, \gamma\right) \triangleq 0$.

**Recursions:** For $k = 1, 2, \ldots, n$, generate the optimal $k$-stage reward-to-go functions as

$$J^{(k)}(\underline{m}_{t-1}, \mathbf{K}_{t-1}, \gamma) = \max_{\{\underline{d}_t, e_t, g_t\}} \Big\{\Omega\left(\underline{m}_{t-1}, \mathbf{K}_{t-1}, \underline{d}_t, e_t, g_t, \gamma\right)$$
$$+ \mathrm{E}\left[J^{(k-1)}(\underline{m}_t, \mathbf{K}_t, \gamma)\right]\Big\}. \quad (108)$$

At the same time, we obtain the optimal $k$th-stage policy $P^{(k)}\left(\underline{s}_t \mid \underline{s}_{t-1}, \alpha_{t-1}(\cdot)\right)$ as defined by the following source coefficients

$$\left\{\underline{d}^{(k)}(\underline{m}_{t-1}, \mathbf{K}_{t-1}, \gamma), e^{(k)}(\underline{m}_{t-1}, \mathbf{K}_{t-1}, \gamma),\right.$$
$$\left. g^{(k)}(\underline{m}_{t-1}, \mathbf{K}_{t-1}, \gamma)\right\}$$
$$\triangleq \arg \max_{\{\underline{d}_t, e_t, g_t\}} \left\{ \Omega\left(\underline{m}_{t-1}, \mathbf{K}_{t-1}, \underline{d}_t, e_t, g_t, \gamma\right) \right.$$
$$\left. + \mathrm{E}\left[J^{(k-1)}(\underline{m}_t, \mathbf{K}_t, \gamma)\right] \right\}. \quad (109)$$

Here, in (108) and (109), the terms $\underline{m}_t$ and $\mathbf{K}_t$ are computed by the Kalman-Bucy equations (106) and (107), respectively.

**Optimized source:** The optimal Gauss-Markov source distribution is

$$\mathcal{P}_\alpha^{\mathrm{GM}} = \left\{ P_t\left(\underline{s}_t \mid \underline{s}_{t-1}, \alpha_{t-1}(\cdot)\right) = P^{(n-t+1)}\left(\underline{s}_t \mid \underline{s}_{t-1}, \alpha_{t-1}(\cdot)\right), \right.$$
$$\left. t = 1, 2, \ldots, n \right\}, \quad (110)$$

where $P^{(n-t+1)}\left(\underline{s}_t \mid \underline{s}_{t-1}, \alpha_{t-1}(\cdot)\right)$ is the optimal $(n-t+1)$th-stage policy obtained by running the above iterations.

**End.**

---

The above value iteration algorithm determines the source distribution $\mathcal{P}_\alpha^{\mathrm{GM}}$ that maximizes the information $I\left(\mathcal{M}; Y_1^n \mid \underline{S}_0 = \underline{s}_0\right)$ for any given shadow price $\gamma > 0$ (or the corresponding finite power budget $P$), see [23] (Ch.7 and Volume II). Each stage of the value iteration determines one set of Gauss-Markov source coefficients used for one transmission as in (110). Notice that the source distribution $\mathcal{P}_\alpha^{\mathrm{GM}}$ in Corollary 4.1 (or equivalently the coefficients $\underline{d}_t$ and $e_t$ in Lemma 1) has a dependence on the conditional mean $\underline{m}_{t-1}$. We next show that this dependence on $\underline{m}_{t-1}$ can be dropped, and the optimal signal can be characterized more explicitly.

*Theorem 5:* There exists a feedback-capacity-achieving feedback-dependent Gauss-Markov source distribution $\mathcal{P}_\alpha^{\mathrm{GM}}$ of the kind as given in Lemma 1, whose coefficients have the following form

$$\underline{d}_t = \underline{d}_t\left(\underline{m}_{t-1}, \mathbf{K}_{t-1}\right) = \underline{d}_t\left(\mathbf{K}_{t-1}\right), \quad (111)$$
$$e_t = e_t\left(\underline{m}_{t-1}, \mathbf{K}_{t-1}\right) = e_t\left(\mathbf{K}_{t-1}\right), \quad (112)$$
$$g_t = g_t\left(\underline{m}_{t-1}, \mathbf{K}_{t-1}\right) = -\left(\underline{d}_t\left(\mathbf{K}_{t-1}\right)\right)^{\mathrm{T}} \underline{m}_{t-1}. \quad (113)$$

That is, an input signal of the form

$$X_t = \underline{d}_t^{\mathrm{T}} \underline{S}_{t-1} + e_t Z_t + g_t = \underline{d}_t^{\mathrm{T}}(\underline{S}_{t-1} - \underline{m}_{t-1}) + e_t Z_t, (114)$$

achieves the $n$-block feedback channel capacity $C^{\mathrm{fb}(n)}$. Further, the processes $\mathbf{K}_t$, $\underline{d}_t$ and $e_t$ are all deterministic and can be determined off-line before the transmission starts. □

*Proof:* It suffices to prove that, for any $k > 0$, the $k$th-stage optimal policy obtained from the value iteration algorithm (108) and (109) has the following structure

$$\underline{d}^{(k)}\left(\underline{m}_{t-1}, \mathbf{K}_{t-1}, \gamma\right) = \underline{d}^{(k)}\left(\mathbf{K}_{t-1}, \gamma\right), \quad (115)$$
$$e^{(k)}\left(\underline{m}_{t-1}, \mathbf{K}_{t-1}, \gamma\right) = e^{(k)}\left(\mathbf{K}_{t-1}, \gamma\right), \quad (116)$$
$$g^{(k)}\left(\underline{m}_{t-1}, \mathbf{K}_{t-1}, \gamma\right) = -\left(\underline{d}^{(k)}\left(\mathbf{K}_{t-1}, \gamma\right)\right)^{\mathrm{T}} \underline{m}_{t-1}. (117)$$

We show by induction on $k$ that the $k$-stage reward-to-go function is independent of the realization of the mean $\underline{m}_{t-1}$, i.e.,

$$J^{(k)}\left(\underline{m}_{t-1}, \mathbf{K}_{t-1}, \gamma\right) = J^{(k)}\left(\mathbf{K}_{t-1}, \gamma\right), \quad (118)$$

and (115), (116) and (117) will be the byproducts. Equality (118) is trivially true for $k = 0$ since $J^{(0)}\left(\underline{m}_{t-1}, \mathbf{K}_{t-1}, \gamma\right) = 0$ by definition. Now, let us assume that (118) is true for $k-1$, where $k \geq 1$, i.e.,

$$J^{(k-1)}\left(\underline{m}_{t-1}, \mathbf{K}_{t-1}, \gamma\right) = J^{(k-1)}\left(\mathbf{K}_{t-1}, \gamma\right). \quad (119)$$

By utilizing the inductive assumption (119), we can rewrite the value iteration (108) as

$$J^{(k)}(\underline{m}_{t-1}, \mathbf{K}_{t-1}, \gamma)$$
$$= \max_{\{\underline{d}_t, e_t, g_t\}} \left\{ \Omega\left(\underline{m}_{t-1}, \mathbf{K}_{t-1}, \underline{d}_t, e_t, g_t, \gamma\right) \right.$$
$$\left. + \mathrm{E}\left[J^{(k-1)}\left(\mathbf{K}_t, \gamma\right)\right] \right\} \quad (120)$$
$$= \max_{\{\underline{d}_t, e_t, g_t\}} \left\{ \Omega\left(\underline{m}_{t-1}, \mathbf{K}_{t-1}, \underline{d}_t, e_t, g_t, \gamma\right) \right.$$
$$\left. + J^{(k-1)}\left(\mathbf{K}_t, \gamma\right) \right\}. \quad (121)$$

Here, we drop the expectation operator in (121) because the conditional covariance matrix $\mathbf{K}_t = F_{KB}^{(\mathbf{K})}\left(\mathbf{K}_{t-1}, \underline{d}_t, e_t\right)$ is a (*deterministic*) function of $\mathbf{K}_{t-1}$, $\underline{d}_t$ and $e_t$. We also note that in order to achieve the maximization in (121), we need to have

$$g_t = -\underline{d}_t^{\mathrm{T}} \underline{m}_{t-1}. \quad (122)$$

This is obvious by noting that $J^{(k-1)}\left(\mathbf{K}_t, \gamma\right) = J^{(k-1)}\left(F_{\mathrm{KB}}^{(\mathbf{K})}\left(\mathbf{K}_{t-1}, \underline{d}_t, e_t\right), \gamma\right)$ is independent of the source coefficient $g_t$, and that, for any choice of coefficients $\underline{d}_t$ and $e_t$, the reward function $\Omega\left(\underline{m}_{t-1}, \mathbf{K}_{t-1}, \underline{d}_t, e_t, g_t, \gamma\right)$ defined in Definition 2 is maximized by $g_t = -\underline{d}_t^{\mathrm{T}} \underline{m}_{t-1}$.

Now, we only need to show that for any two different realizations $\underline{m}_{t-1}$ and $\underline{\tilde{m}}_{t-1}$ of the conditional mean of the channel state, where $\underline{m}_{t-1} \neq \underline{\tilde{m}}_{t-1}$, if $(\underline{d}_t, e_t, g_t) = \left(\underline{d}_t, e_t, -\underline{d}_t^{\mathrm{T}} \underline{m}_{t-1}\right)$ is optimal for the pair $(\underline{m}_{t-1}, \mathbf{K}_{t-1})$, then $(\underline{d}_t, e_t, \tilde{g}_t) = \left(\underline{d}_t, e_t, -\underline{d}_t^{\mathrm{T}} \underline{\tilde{m}}_{t-1}\right)$ must be optimal for $(\underline{\tilde{m}}_{t-1}, \mathbf{K}_{t-1})$, and that

$$J^{(k)}\left(\underline{m}_{t-1}, \mathbf{K}_{t-1}, \gamma\right) = J^{(k)}\left(\underline{\tilde{m}}_{t-1}, \mathbf{K}_{t-1}, \gamma\right). \quad (123)$$

These are verified by checking, for any choice of $\underline{d}_t$ and $e_t$, the following equality

$$\Omega\left(\underline{m}_{t-1}, \mathbf{K}_{t-1}, \underline{d}_t, e_t, -\underline{d}_t^{\mathrm{T}} \underline{m}_{t-1}, \gamma\right)$$
$$+ J^{(k-1)}\left(F_{\mathrm{KB}}^{(\mathbf{K})}\left(\mathbf{K}_{t-1}, \underline{d}_t, e_t\right), \gamma\right)$$
$$= \Omega\left(\underline{\tilde{m}}_{t-1}, \mathbf{K}_{t-1}, \underline{d}_t, e_t, -\underline{d}_t^{\mathrm{T}} \underline{\tilde{m}}_{t-1}, \gamma\right)$$
$$+ J^{(k-1)}\left(F_{\mathrm{KB}}^{(\mathbf{K})}\left(\mathbf{K}_{t-1}, \underline{d}_t, e_t\right), \gamma\right), (124)$$

which is obvious from the Definition 2 for $\Omega(\cdot)$ and the inductive assumption (119). We have thus verified that (118) is true for every $k \geq 0$, and as a direct consequence, (115) and (116) follow. The equality (117) is validated by substituting (115) into equation (122). Finally, (111)-(113) follow from (115)-(117) by the optimality of the value iteration algorithm.

The covariance matrix sequence $\mathbf{K}_t$ can be computed recursively by Kalman-Bucy filtering, i.e., $\mathbf{K}_t = F_{\mathrm{KB}}^{(\mathbf{K})}\left(\mathbf{K}_{t-1}, \underline{d}_t\left(\mathbf{K}_{t-1}\right), e_t\left(\mathbf{K}_{t-1}\right)\right)$, which does not depend on the realization of the random channel output $y_t$. Thus, for any $\mathbf{K}_0$, the sequence $\mathbf{K}_t$ is deterministic, and so are $\underline{d}_t = \underline{d}_t\left(\mathbf{K}_{t-1}\right)$ and $e_t = e_t\left(\mathbf{K}_{t-1}\right)$. ∎

As shown in Theorem 5, in order to achieve the $n$-block feedback capacity $C^{\mathrm{fb}(n)}$, we only need to consider a very simple channel input signal of the following form

$$X_t = \underline{d}_t^{\mathrm{T}} \underline{S}_{t-1} + e_t Z_t + g_t = \underbrace{\underline{d}_t^{\mathrm{T}}(\underline{S}_{t-1} - \underline{m}_{t-1})}_{\text{Kalman innovation}} + e_t Z_t, \quad (125)$$

and choose the parameters $\underline{d}_t$ and $e_t$ properly. Such a feedback-capacity-achieving signal characterization in Theorem 5 also asserts that the center-of-gravity encoding rule (formulated in [16] for memoryless channels) is also optimal for channels with memory. The expected value of the input signal at the receiver's site should always be zero, i.e.,

$$\mathrm{E}\left[X_t \,\middle|\, y_1^{t-1}, \underline{s}_0\right] = 0. \quad (126)$$

An intuitive explanation is that the mean of the signal, i.e., $\mathrm{E}\left[X_t \,\middle|\, y_1^{t-1}, \underline{s}_0\right]$, is always known to both the transmitter and the receiver (i.e., it is deterministic) and thus it does not carry any useful information but only wastes energy. Clearly this waste (the mean) should be equated to zero. The amount of information that is carried by each symbol $X_t$ is determined by the conditional variance of the channel input.

The optimal linear signaling in Theorem 5 also conforms with the linear characterization shown by Cover and Pombra [7], see (14). We note that our new Kalman-Bucy filtering structure admits an encoder whose complexity does not grow with time, and the above filtering structure gives rise to a computation algorithm that breaks the $n$-block feedback capacity computation problem into $n$ sequential stages, where in each stage only $O(L^2)$ variables need to be optimized (since the dimension of the matrix $K_t$ is $L \times L$).

It is interesting to note that the linear signaling form in (125) has already been used as a code by Butman for autoregressive (AR) Gaussian noise channels, see (9)-(11), and equations (1) and (28) in [11], where Butman assumed that $e_t = 0$ for $t > 1$, but provided no proof that such a code is optimal. Butman also showed that the Kalman-Bucy filter needs to be utilized with his chosen code, see equation (29) in [11]. While we still cannot confirm that Butman's choice of parameters $e_t = 0$ for $t > 0$ is optimal, we can now confirm that Butman's code, at least parametrically, matches the optimal solution for AR Gaussian noise channels.

By Theorem 5, the source coefficients $\underline{d}_t$ and $e_t$ and the conditional covariance matrix $\mathbf{K}_t$ are all *deterministic* and can be computed off-line. Thus, the stochastic dynamic programming Algorithm 1 can be simplified, and we obtain the following *deterministic* dynamic programming Algorithm 2, which can be executed off-line (without actually transmitting any signals through the channel). Note that Algorithm 1 on the other hand is *not deterministic* because the expectations in (108) and (109) need to be computed (typically by Monte Carlo methods).

**Algorithm 2** FOR OPTIMIZING THE FEEDBACK-CAPACITY-ACHIEVING SOURCE DISTRIBUTION FOR ANY GIVEN SHADOW PRICE $\gamma > 0$

**Initialization:** For any possible value of the covariance matrix $\mathbf{K}_{t-1}$, define the *terminal reward-to-go function* as $J^{(0)}\left(\mathbf{K}_{t-1}, \gamma\right) = 0$.

**Recursions:** For $k = 1, 2, \ldots, n$, generate the optimal $k$-stage reward-to-go functions as

$$J^{(k)}(\mathbf{K}_{t-1}, \gamma) = \max_{\{\underline{d}_t, e_t\}} \Big\{ \Omega\Big(\underline{m}_{t-1}, \mathbf{K}_{t-1}, \underline{d}_t, e_t, -\underline{d}_t^{\mathrm{T}} \underline{m}_{t-1}, \gamma\Big) + J^{(k-1)}(\mathbf{K}_t, \gamma) \Big\}, \quad (127)$$

and obtain the optimal $k$th-stage policy $P^{(k)}\left(\underline{s}_t \,\middle|\, \underline{s}_{t-1}, \alpha_{t-1}(\cdot)\right)$ as defined by

$$\begin{aligned}&\left\{\underline{d}^{(k)}(\mathbf{K}_{t-1}, \gamma), e^{(k)}(\mathbf{K}_{t-1}, \gamma)\right\} \\&\stackrel{\triangle}{=} \arg \max_{\{\underline{d}_t, e_t\}} \Big\{ \Omega\Big(\underline{m}_{t-1}, \mathbf{K}_{t-1}, \underline{d}_t, e_t, -\underline{d}_t^{\mathrm{T}} \underline{m}_{t-1}, \gamma\Big) \\&\qquad + J^{(k-1)}(\mathbf{K}_t, \gamma) \Big\}. \quad (128)\end{aligned}$$

Here, in (127) and (128), the terms $\underline{m}_t$ and $\mathbf{K}_t$ are computed by the Kalman-Bucy equations (106) and (107), respectively.

**Optimized source:** The optimal Gauss-Markov source distribution is

$$\mathcal{P}_\alpha^{\mathrm{GM}} = \Big\{ P_t\big(\underline{s}_t \big| \underline{s}_{t-1}, \alpha_{t-1}(\cdot)\big) = P^{(n-t+1)}\big(\underline{s}_t \big| \underline{s}_{t-1}, \alpha_{t-1}(\cdot)\big), \\ t = 1, 2, \ldots, n \Big\}, \quad (129)$$

where $P^{(n-t+1)}\left(\underline{s}_t \,\middle|\, \underline{s}_{t-1}, \alpha_{t-1}(\cdot)\right)$ is the optimal $(n-t+1)$th-stage policy.

**End.**

After the source coefficients $\underline{d}_t$ and $e_t$ are optimized by running Algorithm 2, by combining Theorem 4, Lemma 2 and Theorem 5, the power budget $P$ is determined as

$$\begin{aligned}P &= \frac{1}{n} \sum_{t=1}^n \mathrm{E}\left[(X_t)^2 \,\middle|\, \underline{S}_0 = \underline{s}_0\right] \\&= \frac{1}{n} \sum_{t=1}^n \left(\underline{d}_t^{\mathrm{T}} \mathbf{K}_{t-1} \underline{d}_t + (e_t)^2\right), \quad (130)\end{aligned}$$

and the $n$-block feedback channel capacity $C^{\mathrm{fb}(n)}$ can be determined as

$$\begin{aligned}C^{\mathrm{fb}(n)} &= \frac{1}{n} \sum_{t=1}^n \left[h\left(Y_t \,\middle|\, Y_1^{t-1}, \underline{s}_0\right) - \frac{1}{2} \log\left(2\pi e \sigma_W^2\right)\right] \\&= \frac{1}{n} \sum_{t=1}^n \frac{1}{2} \log\left(\frac{\sigma_W^2 + (\underline{a} + \underline{c} + \underline{d}_t)^{\mathrm{T}} \mathbf{K}_{t-1}(\underline{a} + \underline{c} + \underline{d}_t) + (e_t)^2}{\sigma_W^2}\right), \quad (131)\end{aligned}$$

both of which can be computed off-line (without actually transmitting any symbols). We note that the $n$-block feedback capacity (131) is independent of the initial channel state $\underline{s}_0$.

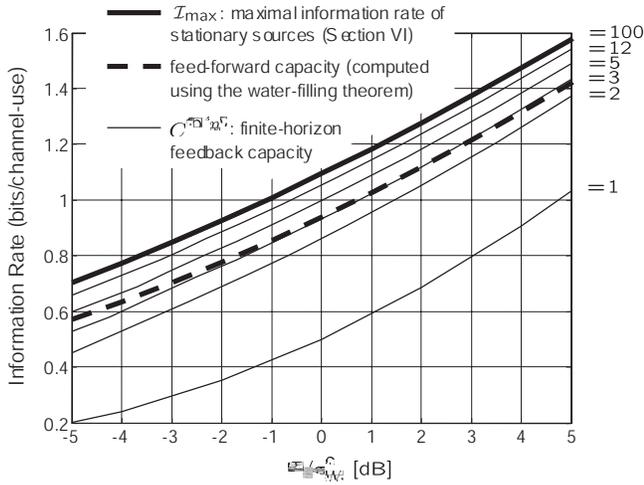

Fig. 8. The feedback capacity for a first-order linear Gaussian noise channel ($a = 0.5$, $c = 0.95$)

### E. Complexity Analysis of Algorithm 2 and Feedback Capacity Curves

Directly computing the $n$-block feedback capacity using the formula in [7], see (13), involves $O(n^2)$ unknown variables. Here, we briefly analyze the complexity of the dynamic programming Algorithm 2. In Algorithm 2, the unknown variables are the length-$L$ vector $\underline{d}_t$, the scalar $e_t$ and the $L \times L$ covariance matrix $\mathbf{K}_t$ for $t = 1, 2, \ldots, n$. The total number of free parameters is thus $\kappa n$, where $\kappa = \frac{L^2+L}{2} + 1$ and $L$ is the channel memory order. Each iteration in Algorithm 2 in general requires an expensive exhaustive search (which could be implemented by appropriately quantizing the search space). For reasonable channel memory orders $L$, reasonable block lengths $n$ and reasonable numerical accuracies (as determined by the quantization step), Algorithm 2 can be easily carried out by ordinary computers. We have plotted the results of such a computation in Figure 8 where the computation takes several minutes. The figure shows that when the block length $n$ increases, the $n$-block feedback capacity very quickly saturates, such that the capacity curves $C^{\text{fb}(n)}$ for $n > 15$ reach saturation. The figure also shows the maximal feedback information rate $\mathcal{I}_{\max}$ of stationary sources, which we derive in Section VI. We also note that for large $n$ the curves $C^{\text{fb}(n)}$ seem to be numerically indistinguishable from $\mathcal{I}_{\max}$, suggesting that for large $n$ the $n$-block feedback capacities $C^{\text{fb}(n)}$ converge to the maximum information rate $\mathcal{I}_{\max}$ achieved by stationary sources.

It is interesting to contrast the convex programming approach of [19] to our approach (Algorithm 2) in terms of complexity and accuracy. In the convex optimization approach of [19], the number of variables is proportional to $n^2$, but the method gives provable complexity vs. accuracy bounds. In our Algorithm 2, the number of variables is linearly proportional to $n$, but the numerical accuracy and effectiveness of Algorithm 2 depend on how we quantize the search space, i.e., how we quantize the possible values of parameters $\mathbf{K}_{t-1}$, $\underline{m}_{t-1}$, $d_t$ and $e_t$. We can efficiently run Algorithm 2 by choosing a good reference point for determining the quantization range of the parameters. One good reference point is the optimal *stationary* source that achieves $\mathcal{I}_{\max}$. In the next section (Section VI), we develop a theory for explicitly determining the optimal stationary feedback source (or feedback code) and the corresponding maximal information rate $\mathcal{I}_{\max}$.

## VI. THE MAXIMAL FEEDBACK INFORMATION RATE OF STATIONARY SOURCES

So far, we have derived a simple linear signaling and feedback strategy that can achieve the $n$-block feedback capacity for any finite horizon depth $n$. We have also given a simple dynamic programming Algorithm 2 to optimize the signal and thus compute the $n$-block feedback capacity. We now concentrate our attention on (asymptotically) stationary sources and their feedback information rates corresponding to $n \to \infty$.

### A. Maximal Information Rates Achieved by Stationary Feedback-Dependent Sources

As shown in Sections IV and V, any achievable information rate can be reached by a feedback-dependent Gauss-Markov (not necessarily stationary) source. Thus, we only need to consider feedback-dependent Gauss-Markov sources in the form shown in Theorem 5. We first define the stationary feedback-dependent Gauss-Markov sources that correspond to the steady state of the Kalman-Bucy feedback filter.

*Definition 3:* [*Stationary feedback-dependent Gauss-Markov sources*] A stationary feedback-dependent Gauss-Markov source is a source that induces stationary channel input and output processes.

An *asymptotically* stationary feedback-dependent Gauss-Markov source, in its limit as $t \to \infty$, induces stationary channel input and output processes. □

*Lemma 3:* For a stationary (or an asymptotically stationary) feedback-dependent Gauss-Markov source, the Kalman-Bucy covariance matrix $\mathbf{K}_t$ and source coefficients $\underline{d}_t$ and $e_t$ converge, i.e.,

$$\lim_{t \to \infty} \mathbf{K}_t = \mathbf{K}, \qquad (132)$$
$$\lim_{t \to \infty} \underline{d}_t = \underline{d}, \qquad (133)$$
$$\lim_{t \to \infty} e_t = e. \qquad (134)$$

Here, the matrix $\mathbf{K}$ satisfies the stationary Kalman-Bucy filter equation (the algebraic Riccati equation)

$$\begin{aligned}\mathbf{K} &= F_{KB}^{(\mathbf{K})}(\mathbf{K}, \underline{d}, e) \\ &= \mathbf{Q}\mathbf{K}\mathbf{Q}^{\text{T}} + \underline{b}\,\underline{b}^{\text{T}} e^2 \\ &\quad - \frac{\left(\mathbf{Q}\mathbf{K}(\underline{a}+\underline{c}+\underline{d})+\underline{b}\,e^2\right)\left(\mathbf{Q}\mathbf{K}(\underline{a}+\underline{c}+\underline{d})+\underline{b}\,e^2\right)^{\text{T}}}{(\underline{a}+\underline{c}+\underline{d})^{\text{T}}\mathbf{K}(\underline{a}+\underline{c}+\underline{d})+e^2+\sigma_W^2},\end{aligned} \quad (135)$$

where $\mathbf{Q} = \mathbf{A} + \underline{b}\,\underline{d}^{\text{T}}(\mathbf{K})$. □

*Proof:* Since the stationary (or asymptotically stationary) source induces, in its limit as $t \to \infty$, stationary channel input and output processes, the Kalman-Bucy filter has a steady state, and thus the sequences $\mathbf{K}_t$, $\underline{d}_t$ and $e_t$ converge.

Equation (135) is then obtained as the stationary form of the covariance matrix propagation equation (107) of the Kalman-Bucy filter. ∎

*Theorem 6:* [*Maximal Information Rates for Stationary Feedback-dependent Sources*] For a power constrained linear Gaussian noise channel depicted in Figure 3 whose noise has a rational power spectral density

$$S_N(\omega) = \sigma_W^2 \left|H(e^{j\omega})\right|^2$$
$$= \sigma_W^2 \frac{\left(1 - \sum_{l=1}^{L} a_l e^{-jl\omega}\right)\left(1 - \sum_{l=1}^{L} a_l e^{jl\omega}\right)}{\left(1 + \sum_{l=1}^{L} c_l e^{-jl\omega}\right)\left(1 + \sum_{l=1}^{L} c_l e^{jl\omega}\right)}, \quad (136)$$

the maximal information rate achieved by (asymptotically) stationary feedback-dependent sources subject to the average input power constraint

$$\lim_{n \to \infty} \frac{1}{n} \sum_{t=1}^{n} \mathrm{E}\left[(X_t)^2 \,|\, \underline{S}_0 = \underline{s}_0\right] \leq P, \quad (137)$$

equals

$$\mathcal{I}_{\max} = \max_{\underline{d}, e} \frac{1}{2} \log\left(\frac{\sigma_W^2 + (\underline{a} + \underline{c} + \underline{d})^{\mathrm{T}} \mathbf{K} (\underline{a} + \underline{c} + \underline{d}) + e^2}{\sigma_W^2}\right), (138)$$

where the maximization in (138) is taken under the following two constraints

$$\underline{d}^{\mathrm{T}} \mathbf{K} \underline{d} + e^2 = P, \quad (139)$$
$$\mathbf{K} = \mathbf{Q} \mathbf{K} \mathbf{Q}^{\mathrm{T}} + \underline{b}\,\underline{b}^{\mathrm{T}} e^2$$
$$- \frac{\left(\mathbf{Q}\mathbf{K}(\underline{a}+\underline{c}+\underline{d}) + \underline{b}\,e^2\right)\left(\mathbf{Q}\mathbf{K}(\underline{a}+\underline{c}+\underline{d}) + \underline{b}\,e^2\right)^{\mathrm{T}}}{(\underline{a}+\underline{c}+\underline{d})^{\mathrm{T}} \mathbf{K} (\underline{a}+\underline{c}+\underline{d}) + e^2 + \sigma_W^2}. \quad (140)$$

Here, $\underline{a} \triangleq [a_1, a_2, \cdots, a_L]^{\mathrm{T}}$, $\underline{c} \triangleq [c_1, c_2, \cdots, c_L]^{\mathrm{T}}$, and matrix $\mathbf{A}$ and vector $\underline{b}$ are

$$\mathbf{A} \triangleq \begin{bmatrix} a_1 & a_2 & \ldots & a_{L-1} & a_L \\ 1 & 0 & \ldots & 0 & 0 \\ 0 & 1 & \ldots & 0 & 0 \\ \vdots & \vdots & \ddots & \vdots & \vdots \\ 0 & 0 & \ldots & 1 & 0 \end{bmatrix}, \quad \underline{b} \triangleq \begin{bmatrix} 1 \\ 0 \\ 0 \\ \vdots \\ 0 \end{bmatrix}.$$

The matrix $\mathbf{Q}$ is defined as $\mathbf{Q} \triangleq \mathbf{A} + \underline{b}\,\underline{d}^{\mathrm{T}}$, and the matrix $\mathbf{K}$ is constrained to be non-negative definite. □

*Proof:* By Lemma 3, for any (asymptotically) stationary Gauss-Markov source, the sequences $\mathbf{K}_t$, $\underline{d}_t$ and $e_t$ converge as $t \to \infty$, so we have

$$\lim_{t \to \infty} \frac{1}{2} \log\left(\frac{\sigma_W^2 + (\underline{a}+\underline{c}+\underline{d}_t)^{\mathrm{T}} \mathbf{K}_{t-1}(\underline{a}+\underline{c}+\underline{d}_t) + (e_t)^2}{\sigma_W^2}\right)$$
$$= \frac{1}{2} \log\left(\frac{\sigma_W^2 + (\underline{a}+\underline{c}+\underline{d})^{\mathrm{T}} \mathbf{K}(\underline{a}+\underline{c}+\underline{d}) + e^2}{\sigma_W^2}\right). \quad (141)$$

Combining Lemma 2 with Theorem 5 and Lemma 3, we also get

$$\lim_{t \to \infty} \mathrm{E}\left[(X_t)^2 \,|\, \underline{S}_0 = \underline{s}_0, Y_1^{t-1} = y_1^{t-1}\right] = \lim_{t \to \infty} \left(\underline{d}_t^{\mathrm{T}} \mathbf{K}_{t-1} \underline{d}_t + (e_t)^2\right)$$
$$= \underline{d}^{\mathrm{T}} \mathbf{K} \underline{d} + e^2. \quad (142)$$

The information rate and the average channel input power can now be easily computed as the Cesáro means of the converging sequences in (141) and (142). From (141) and using equation (31) and Lemma 2, the information rate for the (asymptotically) stationary source exists and equals

$$\lim_{n \to \infty} \frac{1}{n} I\left(\mathcal{M}; Y_1^n \,|\, \underline{S}_0 = \underline{s}_0\right)$$
$$= \lim_{n \to \infty} \frac{1}{2n} \sum_{t=1}^{n} \log\left(\frac{\sigma_W^2 + (\underline{a}+\underline{c}+\underline{d}_t)^{\mathrm{T}} \mathbf{K}_{t-1}(\underline{a}+\underline{c}+\underline{d}_t) + (e_t)^2}{\sigma_W^2}\right) \quad (143)$$
$$= \frac{1}{2} \log\left(\frac{\sigma_W^2 + (\underline{a}+\underline{c}+\underline{d})^{\mathrm{T}} \mathbf{K}(\underline{a}+\underline{c}+\underline{d}) + e^2}{\sigma_W^2}\right). \quad (144)$$

From (142), the average signal power equals

$$\lim_{n \to \infty} \frac{1}{n} \sum_{t=1}^{n} \mathrm{E}\left[(X_t)^2 \,|\, \underline{S}_0 = \underline{s}_0\right]$$
$$= \lim_{n \to \infty} \frac{1}{n} \mathrm{E}_{Y_1^n}\left[\sum_{t=1}^{n} \mathrm{E}\left[(X_t)^2 \,|\, \underline{S}_0 = \underline{s}_0, Y_1^{t-1} = y_1^{t-1}\right]\right]$$
$$= \lim_{n \to \infty} \frac{1}{n} \sum_{t=1}^{n} \left(\underline{d}_t^{\mathrm{T}} \mathbf{K}_{t-1} \underline{d}_t + (e_t)^2\right) \quad (145)$$
$$= \underline{d}^{\mathrm{T}} \mathbf{K} \underline{d} + e^2. \quad (146)$$

Using the results in (144) and (146), we conclude that the maximal information rate $\mathcal{I}_{\max}$ for (asymptotically) stationary Gauss-Markov sources exists and equals

$$\mathcal{I}_{\max} = \max_{\underline{d}, e} \frac{1}{2} \log\left(\frac{\sigma_W^2 + (\underline{a}+\underline{c}+\underline{d})^{\mathrm{T}} \mathbf{K}(\underline{a}+\underline{c}+\underline{d}) + e^2}{\sigma_W^2}\right), (147)$$

where the maximization in (147) is taken under the power constraint

$$\lim_{n \to \infty} \frac{1}{n} \sum_{t=1}^{n} \mathrm{E}\left[(X_t)^2 \,|\, \underline{S}_0 = \underline{s}_0\right] = \underline{d}^{\mathrm{T}} \mathbf{K} \underline{d} + e^2 \leq P, \quad (148)$$

and also under the constraint that the covariance matrix $\mathbf{K}$ satisfies the stationary Kalman-Bucy equation

$$\mathbf{K} = F_{\mathrm{KB}}^{(\mathbf{K})}\left(\mathbf{K}, \underline{d}, e\right), \quad (149)$$

which is explicitly expressed in (140).

By Corollary 4.1, the inequality in (148) can be substituted by an equality, thus proving the validity of the power constraint (139). ∎

To this end, our attempts to further manipulate the expressions in Theorem 6 into a simpler analytical form have not produced the desired outcome because of the complexity of the Kalman-Bucy (algebraic Riccati) equation (140). However, we can readily find the solution to the non-linear programming problem in Theorem 6 numerically, an example of which is depicted in Figure 9. The asymptotic information rate given by Theorem 6, depicted in Figure 9, is compared to the feed-forward capacity computed by the water-filling method [5], [6], [34].

We note that $\mathcal{I}_{\max}$ is a lower bound on the feedback capacity $C^{\mathrm{fb}}$. Figure 8 shows that $\mathcal{I}_{\max}$ numerically overlaps the $n$-block feedback capacity for long block lengths.

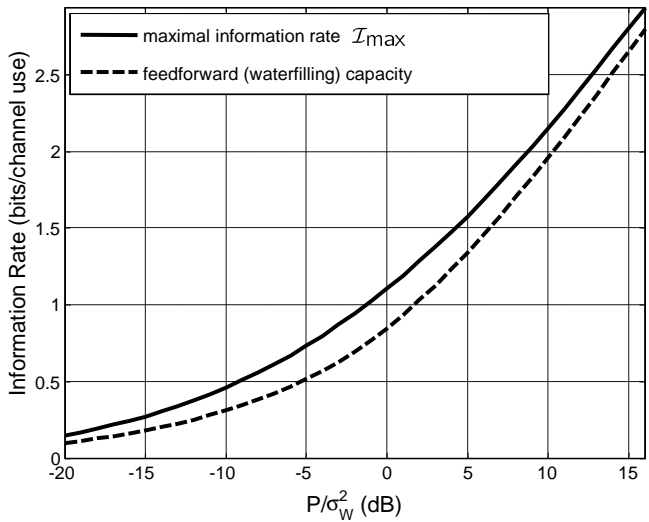

Fig. 9. Maximal information rates achieved by stationary sources over a third order Gaussian noise channel ($\underline{a} = [0, 0.6, 0.4]^T$, $\underline{c} = [0.5, 0.4, 0]^T$)

Under some special circumstances, for example first-order Gaussian channels, we can solve the optimization problem stated in Theorem 6 explicitly. We consider this scenario next.

### B. Maximal Information Rates of Stationary Sources Used over First-Order Channels

We consider the first-order linear-time-invariant (LTI) Gaussian noise channel, where $L = 1$. The channel coefficients $a$ and $c$ (where $-1 < a < 1$ and $-1 < c < 1$) and the channel state covariance $\mathbf{K} = K$ are all scalars. We restrict ourselves to the case $a + c \neq 0$, since otherwise the channel is simply the well studied AWGN channel.

*Theorem 7:* For a first-order Gaussian noise channel characterized by coefficients $a, c$ or equivalently by its noise power spectral density function

$$S_N(\omega) = \sigma_W^2 \frac{(1 - ae^{-j\omega})(1 - ae^{j\omega})}{(1 + ce^{-j\omega})(1 + ce^{j\omega})}, \quad (150)$$

the maximal information rate $\mathcal{I}_{\max}$ achieved by stationary feedback-dependent sources subject to the average input power constraint

$$\lim_{n \to \infty} \frac{1}{n} \sum_{t=1}^n \mathrm{E}\left[(X_t)^2 \,|\, S_0 = s_0\right] \leq P \quad (151)$$

equals

$$\mathcal{I}_{\max} = \frac{1}{2} \log\left(1 + \frac{(1+\eta)^2 P}{\sigma_W^2}\right). \quad (152)$$

Here, the parameter $\eta$ is the largest positive root of the following 4-th order equation

$$\frac{P}{\sigma_W^2}\eta^4 + 2\frac{P}{\sigma_W^2}\eta^3 + \left(\frac{P}{\sigma_W^2} + 1 - a^2\right)\eta^2 \\ -2a(a+c)\eta - (a+c)^2 = 0. \quad (153)$$

The feedback-dependent Gauss-Markov (not necessarily stationary) source that solves (152) has coefficients $d = (a+c)/\eta$ and $e = 0$ in its steady state. □

*Proof:* By Theorem 6, the maximal information rate $\mathcal{I}_{\max}$ is determined by solving the following optimization problem

$$\max_{d,e} \frac{1}{2} \log\left(\frac{\sigma_W^2 + (a+c+d)^2 K + e^2}{\sigma_W^2}\right), \quad (154)$$

under the following constraints

$$d^2 K + e^2 = P, \quad (155)$$

$$K = \frac{(a+d)^2 K \sigma_W^2 + c^2 e^2 K + e^2 \sigma_W^2}{(a+c+d)^2 K + e^2 + \sigma_W^2}. \quad (156)$$

By substituting the constraint (155) into the objective function (154) and noting that the function $\log(\cdot)$ is strictly monotonic, the optimization problem (154), (155) and (156) is equivalent to the following optimization problem

$$\max_{d,e} \left[(a+c+d)^2 K - d^2 K\right], \quad (157)$$

with constraints

$$d^2 K + e^2 = P, \quad (158)$$

$$K^2(a+c+d)^2 + K\left(e^2 + \sigma_W^2 - \sigma_W^2(a+d)^2 - c^2 e^2\right) \\ - e^2 \sigma_W^2 = 0. \quad (159)$$

Obviously, the optimal stationary channel state variance $K$ needs to satisfy $K > 0$.

The Lagrangian function for the optimization problem (157), (158) and (159) is

$$\mathcal{L}(d,e,K,\lambda,\rho) = (a+c+d)^2 K - d^2 K + \lambda(d^2 K + e^2 - P) \\ +\rho \left(K^2(a+c+d)^2 + K\left(e^2 + \sigma_W^2 - \sigma_W^2(a+d)^2 - c^2 e^2\right) \\ - e^2 \sigma_W^2\right), \quad (160)$$

where $\lambda$ and $\rho$ are the Lagrange multipliers. Let the first-order derivatives of the Lagrangian function $\mathcal{L}(d,e,K,\lambda,\rho)$ be zeros, and we have the following necessary conditions for optimality

$$2K\left(a+c+\lambda d + \rho\left(K(a+c+d) - \sigma_W^2(a+d)\right)\right) = 0, \quad (161)$$

$$2e\left[\lambda + \rho\left(K(1-c^2) - \sigma_W^2\right)\right] = 0, \quad (162)$$

$$(a+c+d)^2 - d^2 + \lambda d^2 + \rho\Big(2K(a+c+d)^2 \\ + e^2 + \sigma_W^2 - \sigma_W^2(a+d)^2 - c^2 e^2\Big) = 0, \quad (163)$$

$$d^2 K + e^2 - P = 0, \quad (164)$$

$$K^2(a+c+d)^2 + K\left(e^2 + \sigma_W^2 - \sigma_W^2(a+d)^2 - c^2 e^2\right) \\ - e^2 \sigma_W^2 = 0. \quad (165)$$

We next solve for the optimal values of $d$, $e$ and $K$ from the above equations.

We first prove that $e = 0$ is necessary for optimality. If $e \neq 0$, then equation (162) is substituted by

$$\lambda + \rho\left(K(1-c^2) - \sigma_W^2\right) = 0. \quad (166)$$

We here sketch the proof that (166) cannot hold. If (166) holds, the system of equations (161), (166), (163), (164) and (165) can be solved analytically, and the solution takes one of the following two possible forms. We show that neither of the two forms is acceptable.

1) One possible form of the solution induced by (166) is

$$K = -\frac{-c^2 P + c^2 \sigma_W^2 + 2ac\sigma_W^2}{(a+c)^2 c^2}, \quad (167)$$

$$e^2 = \frac{a^2 \sigma^4 + c^2 P \sigma_W^2}{-c^2 P + c^2 \sigma_W^2 + 2ac\sigma_W^2}. \quad (168)$$

By (167) and (168), the values $K$ and $e^2$ cannot both be positive. However, since $K$ is a variance, we must have $K > 0$, and $e^2$ also must be positive since it is a square of a nonzero real number, so (167) and (168) cannot be the solution.

2) The other possible form of the solution induced by (166) is

$$d = \frac{(1+ac)(a+c)^2 \sigma_W^2}{\sigma_W^2 (a^2 c^3 - 2a^2 c - c - 2a) + P(c^5 - 2c^3 + c)}, \quad (169)$$

$$K = \frac{\sigma_W^2 (a^2 c^3 - 2a^2 c - c - 2a) + P(c^5 - 2c^3 + c)}{(a+c)^2 (c^2 - 1) c}. \quad (170)$$

Now, if we substitute (169) and (170) into the objective function (157), we get a strictly negative value

$$(a+c+d)^2 K - d^2 K$$
$$= \frac{P(c^2 - 1)^2 + \sigma_W^2 (ac - 1)^2}{c^2 - 1} < 0. \quad (171)$$

Since we will compute a positive objective function (157) when the equality condition (166) is replaced by $e = 0$, this negative value (171) cannot be the maximum of (157).

Therefore, we conclude that (166) cannot hold when the variables are optimal and that the proper necessary condition extracted from (162) is $e = 0$.

For $e = 0$, equations (161), (163), (164) and (165) can be solved and the solution takes the following form

$$K = \frac{P}{d^2}, \quad (172)$$

where $d$ satisfies the following 4-th order polynomial equation

$$R(d) = \sigma_W^2 d^4 + 2a\sigma_W^2 d^3 + \left(-P - \sigma_W^2 + a^2 \sigma_W^2\right) d^2$$
$$- 2P(a+c)d - P(a+c)^2$$
$$= 0. \quad (173)$$

Here, we note that the first and last coefficients of $R(d)$ satisfy $\sigma_W^2 > 0$ and $-P(a+c)^2 < 0$. Thus, the polynomial $R(d)$ has either 3 negative and 1 positive real roots or 1 negative and 1 (or 3) positive real roots. By these arguments, we can always select $d$ such that $d(a+c) > 0$ and get a positive objective function to exceed the value in (171), that is

$$(a+c+d)^2 K - d^2 K > 0. \quad (174)$$

Without loss of generality, we let $d$ be

$$d = \frac{a+c}{\eta}. \quad (175)$$

We substitute (175) into (173), and get the 4-th order polynomial equation for $\eta$ as in (153). From the previous discussion, we note that equation (153) always has both positive and negative real roots. We note that $K = P/d^2 = P\eta^2/(a+c)^2$, so the information rate equals

$$\frac{1}{2}\log\left(\frac{\sigma_W^2 + (a+c+d)^2 K}{\sigma_W^2}\right) = \frac{1}{2}\log\left(1 + \frac{(1+\eta)^2 P}{\sigma_W^2}\right), (176)$$

which implies that the optimal value of $\eta$ should be positive to maximize (176). ∎

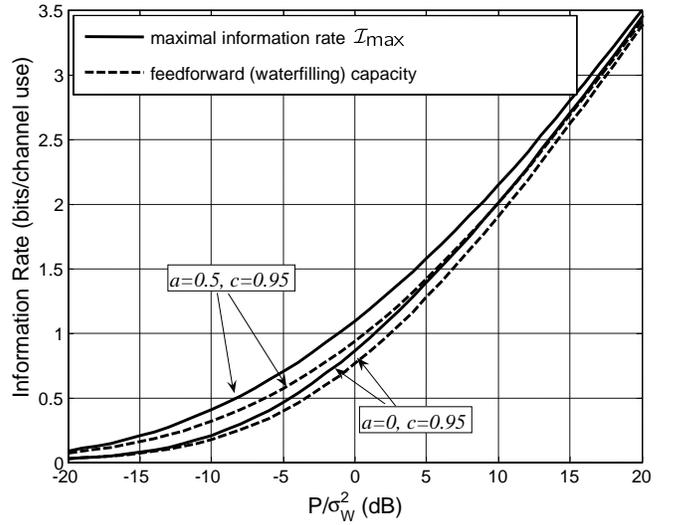

Fig. 10. Maximal information rates for stationary feedback-dependent sources in first-order Gaussian noise channels

In Figure 10, we plot the maximal information rate curves $\mathcal{I}_{\max}$ achieved by (asymptotically) stationary sources for two different first-order LTI Gaussian noise channels (one with first-order AR noise $a = 0$, and the other with first-order ARMA noise). For each channel, we compare the feedback information rate computed by Theorem 7, which is of course the same as computed by Theorem 6, to the feed-forward capacity computed by the water-filling method [5], [6], [34].

An interesting by-product of Theorem 7 is that, with stationary feedback-dependent sources, the maximal information rates for the first-order autoregressive (AR) noise channel (i.e., for $a = 0$ and $c \neq 0$), e.g., the channel $a = 0$ and $c = 0.95$ in Figure 10, equals the well-known Butman feedback capacity lower bound [11]. We next establish a formal proof of this statement.

*Corollary 7.1:* For the first order autoregressive (AR) Gaussian noise channel, where $a = 0$ and $c \neq 0$, the maximal information rate achieved by stationary sources equals

$$\mathcal{I}_{\max} = \frac{1}{2}\log\left(\chi^2\right) = \log\left(|\chi|\right), \quad (177)$$

where the value of $\chi$ satisfies

$$\chi^2 = 1 + \frac{P}{\sigma_W^2}\left(\frac{\chi + |c|}{\chi}\right)^2. \quad (178)$$

This information rate (177) is exactly the same as the achievable rate by Butman's feedback code [11], also given in equation (9). □

*Proof:* By Theorem 7, the optimal stationary source has parameters $e = 0$ and $d = c/\eta$, where $\eta > 0$ in the steady state. The value of $K = P/d^2 = \eta^2 P/c^2$ needs to satisfy the stationary Kalman-Bucy filtering equation (135)

$$\begin{aligned} K &= \frac{\eta^2 P}{c^2} \\ &= F_{\text{KB}}^{(\mathbf{K})}(K, d, e) \\ &= F_{\text{KB}}^{(\mathbf{K})}\left(\frac{\eta^2 P}{c^2}, \frac{c}{\eta}, 0\right) \\ &= \frac{P\sigma_W^2}{\sigma_W^2 + (1+\eta)^2 P}. \end{aligned} \quad (179)$$

From (179), we have

$$\frac{\sigma_W^2 + (1+\eta)^2 P}{\sigma_W^2} = \left(\frac{c}{\eta}\right)^2. \quad (180)$$

Next, we define

$$\chi \triangleq \frac{c}{\eta}. \quad (181)$$

By substituting the definition (181) into equation (180), we get the equation (178) for $\chi$. Further, by substituting (180) into the information rate formula (152), we get

$$\mathcal{I}_{\max} = \frac{1}{2}\log\left(\frac{\sigma_W^2 + (1+\eta)^2 P}{\sigma_W^2}\right) = \log\left(\chi^2\right). \quad (182)$$

∎

Note that Butman's information rate [11] (Corollary 7.1) applies only to autoregressive (AR) noise channels of the first order $L = 1$. The optimality of Butman's code has been a long-standing conjecture (Butman's conjecture was generalized to higher order channels in [18]). Our solution (Theorem 6 and Theorem 7) applies to both autoregressive (AR) and moving-average (MA) (or combined ARMA) noise processes of any finite order $L$. We now have a proof that Butman's code achieves the maximal information rate $\mathcal{I}_{\max}$ of stationary sources for first order AR channels. However, we still cannot claim that this maximal rate $\mathcal{I}_{\max}$ equals the channel capacity $C^{\text{fb}}$. Further, though numerically verifiable, we still cannot *prove* that Butman's code for higher order ($L > 1$) AR channel noise achieves $\mathcal{I}_{\max}$.

### C. Sufficient Condition for the Existence of the Feedback Capacity

We consider the sufficient condition for the limit $C^{\text{fb}} = \lim_{n\to\infty} C^{\text{fb}(n)}$ to exist.

*Definition 4:* [*Time-invariant feedback-dependent Gauss-Markov sources*] A time-invariant feedback-dependent Gauss-Markov source, restricted to the optimal structure shown in Theorem 5, is a source whose coefficients $\underline{d}_t$ and $e_t$ have a time-invariant dependence on the covariance matrix $\mathbf{K}_{t-1}$. The time-dependence is captured by the dependence on the posterior channel state statistics only, i.e.,

$$\underline{d}_t = \underline{d}\left(\mathbf{K}_{t-1}\right), \quad (183)$$
$$e_t = e\left(\mathbf{K}_{t-1}\right), \quad (184)$$
$$g_t = -\underline{d}_t^{\text{T}} \underline{m}_{t-1} = -\underline{d}^{\text{T}}\left(\mathbf{K}_{t-1}\right) \underline{m}_{t-1}. \quad (185)$$

∎

It is interesting to note that a time-invariant source as defined above may in fact induce a channel input process $X_t$ whose parameters $\underline{d}_t$, $e_t$ and the induced process $\mathbf{K}_t$ never reach a steady state as $t \to \infty$. So, time-invariance does not guarantee stationarity[3].

Corollary 4.1 states a finite-horizon stochastic control problem. We next consider the corresponding infinite-horizon problem. In the next lemma, we link the Bellman equation [23] for this infinite-horizon problem to the feedback capacity (should it exist). Subsequently, we formulate a sufficient condition for the feedback capacity $C^{\text{fb}}$ to exist.

*Lemma 4:* Let the function $C(\gamma)$ with argument $\gamma > 0$ be defined as

$$C(\gamma) \triangleq \lim_{n\to\infty} \max_{\mathcal{P}_\alpha^{\text{GM}}} \frac{1}{n}\sum_{t=1}^n \Omega\Big(\underline{m}_{t-1}, \mathbf{K}_{t-1}, \underline{d}_t, e_t,$$
$$g_t = -\underline{d}_t^{\text{T}} \underline{m}_{t-1}, \gamma\Big), \quad (186)$$

if the limit exists. Bellman's equation [23] associated with (186) takes the following form

$$C(\gamma) + \pi(\mathbf{K}, \gamma) = \max_{\underline{d}, e}\left\{\Omega\left(\underline{m}, \mathbf{K}, \underline{d}, e, -\underline{d}^{\text{T}} m, \gamma\right)\right.$$
$$\left. + \pi\left(\tilde{\mathbf{K}}, \gamma\right)\right\}. (187)$$

Here, the function $\pi(\mathbf{K}, \gamma)$ is the *optimal relative reward-to-go* function, the symbol $\tilde{\mathbf{K}}$ on the right-hand side of (187) is the short notation for the Kalman-Bucy filter output, that is

$$\tilde{\mathbf{K}} = F_{KB}^{(\mathbf{K})}\left(\mathbf{K}, \underline{d}, e\right). \quad (188)$$

Bellman's equation (187) is solved by a *time-invariant* source as long as it has a solution. Further, the time-invariant source that solves (187) also solves (186), in which case the asymptotic feedback capacity exists and equals

$$C^{\text{fb}} \triangleq \lim_{n\to\infty} C^{\text{fb}(n)} = C(\gamma) + \gamma P. \quad (189)$$

∎

*Proof:* The lemma is proved by applying the results for the average-cost-per-stage stochastic control problem in [23] (Ch. 7 and Volume II). Note that if there exists a value $C(\gamma)$ and a function $\pi(\mathbf{K}, \gamma)$ which solve Bellman's equation (187), the value $C(\gamma)$ and the corresponding feedback-dependent Gauss-Markov source $\mathcal{P}_\alpha^{\text{GM}}$ determined by coefficients $\underline{d}_t = \underline{d}(\mathbf{K}_{t-1})$ and $e_t = e(\mathbf{K}_{t-1})$ also solves the maximization in (186), and vice versa, see [23] (Ch. 7 and Volume II). The time-invariant source $\mathcal{P}_\alpha^{\text{GM}}$ that solves Bellman's equation (187) is thus optimal for the particular choice of power shadow price $\gamma$.

---

[3]Here, we give an example of a time-invariant but non-stationary source. Suppose that we have a first-order system (such as the one discussed in Section VI-B) for which $a = 0$ and $c = 0.2$. Let the filter coefficients $e_t = 0$ and $d_t$ be time-invariant, but let $d_t$ be dependent on the value of the posterior state variance $K_{t-1}$ as: $d_t = d(K_{t-1}) = -1$ if $K_{t-1} > 0.01$, and $d_t = d(K_{t-1}) = -19$ if $K_{t-1} \leq 0.01$. Note that $d_t$ is indeed a time-invariant function of $K_{t-1}$ because it does not depend on time $t$, but rather on the value of $K_{t-1}$. Let the value of the posterior state variance at time $t - 1$ be $K_{t-1} = 0.0102$. Then, because of this special choice of the parameters, when substituted into the Riccati equation (156) we get $K_t = 0.0062$. Applying the Riccati equation one more time, we get $K_{t+1} = 0.0102 = K_{t-1}$. So, the system oscillates.

From Theorem 4 and Theorem 5, the feedback capacity exists if $C(\gamma)$ defined in (186) exists, and equation (189) follows. ∎

In general, ensuring that Bellman's equation (187) has a solution can be very complicated, see [23] (Ch. 7 and Volume II). However, from Lemma 4, we know that if the solution exists, it must be a time-invariant Gauss-Markov source. One sufficient condition [23] to guarantee the existence of a time-invariant solution to Bellman's equation (187) is that for two arbitrary valid covariance matrices $\bar{\mathbf{K}}$ and $\hat{\mathbf{K}}$, there exists a time-invariant Gauss-Markov source distribution that drives the Kalman-Bucy filter covariance matrix from value $\mathbf{K}_t = \bar{\mathbf{K}}$ to value $\mathbf{K}_{t+\tau} = \hat{\mathbf{K}}$ within finite time $\tau < \infty$. We note that verifying such a sufficient condition is possible only on a case by case basis, and a systematic analytic verification is still missing.

Further, even if the optimal feedback-dependent source is time-invariant, the covariance matrix sequence $\mathbf{K}_t$ may not converge. One way to numerically check whether the sequence $\mathbf{K}_t$ converges for a given channel is to run the dynamic-programming Algorithm 2 (value iteration) or policy iteration [23] for a large block length $n$. However, besides this numerical verification procedure, there are no known systematic approaches to analytically handle such a problem for an arbitrary channel. Therefore, to make further progress towards finding an analytic solution to the feedback capacity problem, we would need to prove that $\mathbf{K}_t$ converges as $t \to \infty$.

By numerically running the dynamic programming Algorithm 2 for various Gaussian noise channels with large block lengths $n$, e.g., $n \geq 100$, we have always observed that the Kalman-Bucy filter in Figure 7 becomes stationary, i.e., $\mathbf{K}_t$ converges numerically, as $t$ becomes large. It has been a long-standing conjecture that stationary sources achieve the feedback capacity [11]. Here, we reformulate the conjecture in terms of the posterior state covariance matrix computed by the Kalman-Bucy filter.

*Conjecture 1:* The optimal (feedback-capacity-achieving) source induces a stationary (or asymptotically stationary) Kalman-Bucy filter for processing the feedback, i.e., for the optimal source, the limit

$$\lim_{t \to \infty} \mathbf{K}_t = \mathbf{K}$$

exists. □

Under Conjecture 1, the feedback capacity of a power constrained linear Gaussian noise would be achieved by a (asymptotically) stationary source, and the feedback capacity would equal $\mathcal{I}_{\max}$ given in Theorem 6.

For first-order moving-average (MA) linear Gaussian noise channels, Kim [35] recently proved that a uniform power allocation over time is asymptotically optimal by following a different approach from what used in this paper, and that the feedback capacity equals the maximal information rate derived in Theorem 7 for the moving-average noise subcase. This new result implies that stationary sources for first-order moving-average (MA) linear Gaussian noise channels are indeed optimal.

## VII. CONCLUSION

We considered the problem of computing the $n$-block feedback capacity of a Gaussian noise channel with memory under an average channel input power constraint. In its full generality, the problem would consider any power spectral density of the Gaussian noise process. However, for technical reasons, we only considered noise processes that have rational power spectra, i.e., noise processes that are either autoregressive (AR) or moving average (MA) or both (ARMA). Since we were computing the capacity of a channel with *memory*, we found it beneficial to cast the problem in the state-space realization formulation, which proved to be well-suited for this problem.

For the Gaussian noise channel with a rational power spectrum, we found that the $n$-block feedback capacity $C^{\text{fb}(n)}$ is achieved by a Gauss-Markov (not necessarily stationary) source distribution, where the channel input depends only on the previous channel state and the posterior channel state distribution computed by a Kalman-Bucy filter. Further, we showed that the channel state, the posterior channel state distribution and the channel output jointly form a Markov process.

The Markov property of the optimal source reduced the $n$-block feedback capacity computation to a standard dynamic-system stochastic control problem, which can be solved by dynamic programming. For this optimization problem, we found a simple structure of the optimal source, where the encoding complexity is constant for any time instant. We showed that the coefficients of the optimal Gauss-Markov source depend only on the covariance matrix of the posterior channel state estimate computed by a Kalman-Bucy filter, and can be optimized deterministically and off-line. The $n$-block optimization problem is thus broken into $n$ sequential problems. In each sequential step, $O(L^2)$ variables need to be solved for, where $L$ is the order of the ARMA channel noise.

We note that for additive white Gaussian noise (AWGN) channels, retransmitting the message uncertainty [15] or transmitting the newly coded signal [1] could both achieve the channel capacity (the feedback capacity equals the feedforward capacity). In our formulation (125), it is still an open problem to determine if both parameters $\underline{d}_t$ and $e_t$ could take non-zero optimal values. For the initial transmission at time $t = 1$, since $\underline{s}_0$ is known, the transmitter has to let $e_t \neq 0$ to start transmission. Our numerical simulation have always suggested that, for $t \to \infty$, the optimal value of $e_t$ should be zero, but a proof is missing.

We solved analytically the maximal feedback information rate $\mathcal{I}_{\max}$ achieved by (asymptotically) stationary sources, which represents a lower bound on the feedback capacity $C^{\text{fb}}$. Under a Kalman-Bucy filter stationarity assumption (Conjecture 1), the feedback capacity $C^{\text{fb}}$ would equal $\mathcal{I}_{\max}$. Conjecture 1 is a reformulation (in terms of Kalman-Bucy filtering parameters) of a long-standing conjecture that stationary sources achieve the feedback capacity.

## APPENDIX I
## ALTERNATIVE PROOF OF THEOREM 2

Here we present an alternative proof of Theorem 2 based on dynamic programming [23] and Lagrange multipliers (see

Appendix 2). Consider the mixed cost:

$$\sum_{t=1}^{n} I(\underline{S}_{t-1}^{t}; Y_t \mid Y_1^{t-1}, \underline{s}_0) - \gamma \cdot \mathrm{E}\left[\sum_{t=1}^{n}(X_t)^2 \mid \underline{s}_0\right]. \quad (190)$$

As discussed in Appendix 2, there is, under an optimal source, a one-to-one relationship between the Lagrange multiplier (shadow price) $\gamma$ and the resulting average power constraint $P$, and hence the optimization of the mixed cost in (190) yields the optimal source. We will now show that for any shadow price $\gamma$, including the $\gamma$ corresponding to the given power $P$, we can without loss of generality restrict ourselves to sources of the form $\{P_t(\underline{s}_t \mid \underline{s}_{t-1}, \alpha_{t-1}(\cdot)), t = 1, 2, \cdots\}$.

As is typical of dynamic programming arguments we will prove this by backward induction starting at time $n$. At time $n$ the optimization is:

$$\max_{P_n(\underline{s}_n \mid \underline{s}_{n-1}, y_1^{n-1})} I(\underline{S}_{n-1}^n; Y_n \mid Y_1^{n-1}, \underline{s}_0) - \gamma \cdot \mathrm{E}[(X_n)^2 \mid \underline{s}_0].$$

This can be rewritten as (191), where $A_{n-1}(\cdot) = P_{\underline{S}_{n-1}\mid\underline{S}_0,Y_1^{n-1}}\left(\cdot \mid \underline{s}_0, Y_1^{n-1}\right)$ is the random function whose realizations are $\alpha_{n-1}(\cdot) = P_{\underline{S}_{n-1}\mid\underline{S}_0,Y_1^{n-1}}\left(\cdot \mid \underline{s}_0, y_1^{n-1}\right)$.

Because $\underline{s}_0$ is fixed and to simplify notation we will not explicitly condition on $\underline{s}_0$ in the following discussion. To compute the first term in the inner expectation of (191), we need access to

$$p(y_n, \underline{s}_{n-1}^n, \alpha_{n-1}(\cdot) \mid y_1^{n-1})$$
$$= p(y_n \mid \underline{s}_{n-1}^n) P_n(\underline{s}_n \mid \underline{s}_{n-1}, y_1^{n-1}) \alpha_{n-1}(\underline{s}_{n-1}).$$

To compute the second term in the inner expectation of (191), we need access to

$$p(x_n \mid y_1^{n-1})$$
$$= \int p(x_n \mid \underline{s}_{n-1}^n) P_n(\underline{s}_n \mid \underline{s}_{n-1}, y_1^{n-1}) \alpha_{n-1}(\underline{s}_{n-1}) d\underline{s}_{n-1}^n.$$

Hence, in the inner expectation of (191), the maximization over the choice of $P_n(\underline{s}_n \mid \underline{s}_{n-1}, y_1^{n-1})$ requires knowing only $\underline{s}_{n-1}$ and $\alpha_{n-1}(\cdot)$. Thus without loss of generality we can restrict the source at time $n$ to be of the form: $P_n(\underline{s}_n \mid \underline{s}_{n-1}, \alpha_{n-1}(\cdot))$. Let the optimal cost-to-go [23] at time $n$, given by the inner expectation in (191), be denoted by $J_n(\alpha_{n-1}(\cdot))$.

Now, via the induction hypothesis assume that the source for times $\tau = t+1, ..., n$ can be chosen without loss of generality to be of the form $\{P_\tau(\underline{s}_\tau \mid \underline{s}_{\tau-1}, \alpha_{\tau-1}(\cdot)), \tau = t+1, ..., n\}$, and assume the cost-to-go functions $J_\tau(\alpha_{\tau-1}(\cdot))$ can be chosen to only depend on $\alpha_{\tau-1}(\cdot)$ for $\tau = t+1, ..., n$. The optimization at time $t$ is given in (192) where $J_{t+1}(\cdot)$ is the optimal cost-to-go (which by the induction hypothesis only depends on $\alpha_t$).

As in (191) we can write (192) as an iterated expectation conditioned on $Y_1^{t-1}$. The inner expectation for the new term $J_{t+1}(\alpha_t(\cdot))$ takes the form

$$\mathrm{E}\left[J_{t+1}(A_t(\cdot)) \mid Y_1^{t-1}\right]. \quad (193)$$

To compute (193), we need access to $p(\alpha_t(\cdot) \mid y_1^{t-1})$. Now by equation (59) we know that $\alpha_t(\cdot)$ is a function of $\alpha_{t-1}(\cdot), y_t,$ and $P_t(\underline{s}_t \mid \underline{s}_{t-1}, y_1^{t-1})$. Therefore, to compute (193), we need access to

$$p(y_t, \alpha_{t-1}(\cdot) \mid y_1^{t-1})$$
$$= \int p(y_t \mid \underline{s}_{t-1}^t) P_t(\underline{s}_t \mid \underline{s}_{t-1}, y_1^{t-1}) \alpha_{t-1}(\underline{s}_{t-1}) d\underline{s}_{t-1}^t.$$

Hence (193) depends on $y_1^{t-1}$ only through $\alpha_{t-1}$ and the choice of source $P_t(\underline{s}_t \mid \underline{s}_{t-1}, y_1^{t-1})$.

This observation, along with an argument similar to the one given for the optimization (191), shows that the optimization in (192) can, without loss of generality, be restricted to a source of the form: $P_t(\underline{s}_t \mid \underline{s}_{t-1}, \alpha_{t-1}(\cdot))$. Thus we have proved Theorem 2.

## APPENDIX II
## STRONG DUALITY AND THE RELATIONSHIP BETWEEN THE POWER CONSTRAINT AND THE SHADOW PRICE

### A. Strong Duality under Slater's Conditions

Let $I(\mathcal{P})$ be a function (not necessarily concave) of a variable $\mathcal{P}$. Let $f(\mathcal{P}) \leq 0$ be a constraint on the variable $\mathcal{P}$. Let

$$\mathcal{P}^* = \arg \max_{\mathcal{P}: f(\mathcal{P}) \leq 0} I(\mathcal{P})$$

be the solution of the constrained optimization problem, and let

$$I^* \triangleq I(\mathcal{P}^*) = \max_{\mathcal{P}: f(\mathcal{P}) \leq 0} I(\mathcal{P})$$

be the constrained maximum. Define the Lagrangian for the constrained optimization problem as

$$\mathcal{L}(\mathcal{P}, \gamma) = I(\mathcal{P}) - \gamma f(\mathcal{P}),$$

for which the Lagrangian dual function is the unconstrained maximum

$$G(\gamma) = \max_{\mathcal{P}}[I(\mathcal{P}) - \gamma f(\mathcal{P})].$$

From Boyd-Vandenberghe [30] Section 5.1.2, we have that $G(\gamma)$ is convex (even when $I(\mathcal{P})$ is not concave), and from Section 5.1.3, we have

$$G(\gamma) \geq I^* = I(\mathcal{P}^*).$$

Let the solution to the Lagrangian dual problem be

$$\gamma^* = \arg \min_\gamma G(\gamma).$$

Weak duality, Boyd-Vandenberghe [30] Section 5.2.2, guarantees

$$G^* \triangleq G(\gamma^*) \geq I(\mathcal{P}^*) = I^*.$$

**Theorem** (Strong duality under Slater's conditions, Boyd-Vandenberghe [30], Section 5.2.3) If $I(\mathcal{P})$ is a concave function, if $f(\mathcal{P})$ is a convex function, and if there exists a parameter $\mathcal{P}$ for which $f(\mathcal{P}) < 0$ is feasible, then strong duality $I^* = G^*$ holds.

$$\mathrm{E}\left[\max_{P_n(\underline{s}_n|\underline{s}_{n-1},y_1^{n-1})} \mathrm{E}\left[\log \frac{p(Y_n|\underline{S}_{n-1}^n)}{\int p(Y_n|\underline{s}_{n-1}^n)P_n(\underline{s}_n|\underline{s}_{n-1},Y_1^{n-1})A_{n-1}(s_{n-1})\mathrm{d}\underline{s}_{n-1}^n} - \gamma(X_n)^2 \bigg| Y_1^{n-1}\right] \bigg| \underline{s}_0\right] \quad (191)$$

$$\max_{P_t(\underline{s}_t|\underline{s}_{t-1},y_1^{t-1})} I(\underline{S}_{t-1}^t;Y_t \mid Y_1^{t-1},\underline{s}_0) - \gamma \cdot \mathrm{E}[(X_t)^2 \mid \underline{s}_0] + \mathrm{E}[J_{t+1}(A_t(\cdot)) \mid \underline{s}_0], \quad (192)$$

### B. Propositions Involving the Feedback Capacity

Let $C^{\mathrm{fb}(n)}(P)$ be the $n$-block feedback capacity under the power constraint $P$, and let $C^{(n)}(P)$ be the feed-forward $n$-block capacity under the same power constraint.

**Proposition 1** (Cover-Pombra [7])

$$C^{\mathrm{fb}(n)}(P) \leq C^{(n)}(P) + \frac{1}{2}.$$

**Proposition 2** (reformulation: water-filling theorem, Gallager [34], Theorem 7.5.1 )

$$C^{(n)}(P) = \frac{1}{2n}\sum_{i=1}^{k}\log\frac{nP+r_1+r_2+\cdots+r_k}{kr_i}$$

where $0 \leq r_1 \leq r_2 \leq \cdots \leq r_n$ are eigenvalues of the $n \times n$ channel noise covariance matrix and $k$ ($\leq n$) is the largest integer satisfying $nP+r_1+r_2+\cdots+r_k > kr_k$.

**Proposition 3**

$$\lim_{P\to\infty}\frac{C^{\mathrm{fb}(n)}(P)}{P} = 0.$$

**proof:** *Using Propositions 1 and 2, we have*

$$\lim_{P\to\infty}\frac{C^{\mathrm{fb}(n)}(P)}{P} \leq \lim_{P\to\infty}\frac{C^{(n)}(P)+\frac{1}{2}}{P}$$

$$= \lim_{P\to\infty}\frac{\frac{1}{2n}\sum_{i=1}^{k}\log\frac{nP+r_1+r_2+\cdots+r_k}{kr_i} + \frac{1}{2}}{P}.$$

*Applying L'Hôpital's rule to the right-hand side proves the proposition.*

**Proposition 4** For any $\gamma > 0$,

$$\lim_{P\to\infty}\left[C^{\mathrm{fb}(n)}(P) - \gamma P\right] = -\infty.$$

**proof:** *Similar to the proof or Proposition 3.*

### C. Strong Duality of the n-Block Feedback Capacity

Define $\mathcal{P}$ as the source (under some proper parametrization; here it is convenient to choose the parametrization in Vandenberghe, Boyd and Wu [19] because it leads to a concave feedback information rate). Let $I^{\mathrm{fb}(n)}(\mathcal{P})$ be the $n$-block feedback information rate achieved by the source $\mathcal{P}$. Let the source $\mathcal{P}$ be subject to a power constraint $\mathrm{P}_{\mathrm{ow}}(\mathcal{P}) \leq P$. The primal constrained optimization problem is then stated as

$$C^{\mathrm{fb}(n)}(P) = \max_{\mathcal{P}:\mathrm{P}_{\mathrm{ow}}(\mathcal{P})\leq P} I^{\mathrm{fb}(n)}(\mathcal{P}).$$

The case $P = 0$ is trivial and can be dismissed. The function $I^{\mathrm{fb}(n)}(\mathcal{P})$ is concave and the power constraint is convex, see Vandenberghe, Boyd and Wu [19]. Further, for any $P > 0$, there exists a feasible source that satisfies $\mathrm{P}_{\mathrm{ow}}(\mathcal{P}) < P$ (just take the trivial zero-source as an example). Hence, Slater's conditions for strong duality are satisfied and the solution of the primal problem equals the solution of the dual problem for any $P > 0$.

The dual problem is formulated as follows. The Lagrangian is

$$\mathcal{L}(\mathcal{P},\gamma) = I^{\mathrm{fb}(n)}(\mathcal{P}) - \gamma\mathrm{P}_{\mathrm{ow}}(\mathcal{P}) + \gamma P.$$

The Lagrangian dual function is the unconstrained maximum

$$G(\gamma) = \max_{\mathcal{P}}\mathcal{L}(\mathcal{P},\gamma).$$

The solution to the dual problem is

$$\gamma^* = \arg\min_{\gamma} G(\gamma),$$

and the $n$-block feedback capacity is

$$C^{\mathrm{fb}(n)}(P) = G^* \stackrel{\triangle}{=} G(\gamma^*) = \min_{\gamma} G(\gamma).$$

### D. Relationship between the Power Constraint and the Shadow Price

Because of the strong duality between the $n$-block feedback capacity computation problem and its dual, for any power constraint $P > 0$, the solution to the dual problem gives a parameter $\gamma^*$ such that $G(\gamma^*)$ is the solution to the primal problem. We now want to establish the backwards relationship, i.e., that for any chosen shadow price $\gamma$, the unconstrained solution to the Lagrangian maximization

$$\mathcal{P}^* = \arg\max_{\mathcal{P}}\left[I^{\mathrm{fb}(n)}(\mathcal{P}) - \gamma\mathrm{P}_{\mathrm{ow}}(\mathcal{P})\right] \quad (194)$$

gives a source $\mathcal{P}^*$ whose power satisfies $\mathrm{P}_{\mathrm{ow}}(\mathcal{P}) = P < \infty$, such that $\mathcal{P}^*$ is the solution to the primal problem when $P$ is the power constraint.

First, we can easily dismiss the shadow price $\gamma = 0$ from consideration because if $\gamma = 0$, then the solution to (194) clearly gives a source whose power is $P = \infty$, i.e., the source is not power-constrained and can be dismissed.

Next, we want to establish that for any power constraint $P < \infty$, there exists an optimal shadow price $\gamma^* > 0$, and vice versa, that for any shadow price $\gamma > 0$, the solution to the unconstrained Lagrangian optimization (194) gives a source whose power $P$ is finite, $P < \infty$.

**Proposition A** For any power constraint $P < \infty$, the solution to the Lagrangian dual problem satisfies $\gamma^* > 0$.

**proof:** Pick any $P < \infty$. Then clearly, $C^{\text{fb}(n)}(P) < \infty$. Because of the strong duality, we must have

$$G(\gamma^*) = C^{\text{fb}(n)}(P) < \infty,$$

where

$$\gamma^* = \arg\min_\gamma G(\gamma).$$

Now, assume that $\gamma^* = 0$. Then the solution to the dual problem delivers $G(\gamma^*) = G(0) = \infty$, which contradicts the previously established relationship $C^{\text{fb}(n)}(P) = G(\gamma^*) < \infty$. Hence, $\gamma^* > 0$ must hold.

**Proposition B** For any shadow price $\gamma > 0$, the solution to the unconstrained Lagrangian optimization problem (194) delivers a source whose power is finite, i.e., $P < \infty$.

**proof:** First notice that

$$\max_{\mathcal{P}} \left[ I^{\text{fb}(n)}(\mathcal{P}) - \gamma \text{P}_{\text{ow}}(\mathcal{P}) \right] \geq 0$$

because we can always pick the trivial all-zero source whose power and information rate are zero. Now, assume that there exists a $\gamma > 0$ such that the solution to (194) delivers a source $\mathcal{P}^*$ whose power is $\text{P}_{\text{ow}}(\mathcal{P}^*) = P = \infty$. For such a source, invoking Proposition 4, the following holds

$$\max_{\mathcal{P}} \left[ I^{\text{fb}(n)}(\mathcal{P}) - \gamma \text{P}_{\text{ow}}(\mathcal{P}) \right]$$
$$= I^{\text{fb}(n)}(\mathcal{P}^*) - \gamma \text{P}_{\text{ow}}(\mathcal{P}^*)$$
$$\leq \lim_{P \to \infty} C^{\text{fb}(n)}(P) - \gamma \text{P}_{\text{ow}}(\mathcal{P}^*)$$
$$= \lim_{P \to \infty} \left[ C^{\text{fb}(n)}(P) - \gamma P \right]$$
$$= -\infty,$$

which contradicts our earlier conclusion. Thus, for any $\gamma > 0$, the solution to (194) must be a source whose power is finite, $P < \infty$.

Propositions A and B jointly establish that any shadow price $\gamma > 0$ maps to a power $P < \infty$, and vice versa. We now show that the solution to the unconstrained Lagrangian optimization (194) gives a source that achieves the feedback capacity for some $P < \infty$.

**Proposition C** For any $\gamma > 0$, the solution to the unconstrained Lagrangian optimization (194) gives a source $\mathcal{P}^*$ with power $\text{P}_{\text{ow}}(\mathcal{P}^*) = P$ for which $\gamma^* = \gamma$ is the solution to the Lagrangian dual optimization.

**proof:** Pick some $\gamma > 0$. For this $\gamma$, Proposition B established that the solution to (194) is a source $\mathcal{P}^*$ whose power $P$ is finite, $P < \infty$. For such finite $P$, strong Lagrangian duality holds, so for this value of $P$, we must have

$$I^{\text{fb}(n)}(\mathcal{P}^*) = C^{\text{fb}(n)}(P) = G(\gamma^*),$$

where $\gamma^*$ is a solution to the Lagrangian dual problem for power $P$. Now, if we substitute this source $\mathcal{P}^*$, whose power is $\text{P}_{\text{ow}}(\mathcal{P}^*) = P$, into the expression for the Lagrangian dual function $G(\cdot)$, for our chosen value $\gamma > 0$, we get

$$G(\gamma) = I^{\text{fb}(n)}(\mathcal{P}^*) = C^{\text{fb}(n)}(P) = G(\gamma^*).$$

Since $G(\cdot)$ is a convex function, and since $G(\gamma^*)$ is the minimum of $G(\cdot)$, it follows that $G(\gamma)$ is also the minimum of $G(\cdot)$. Therefore $\gamma$ is also a solution to the Lagrangian dual problem for power $P$. Hence we can set $\gamma^* = \gamma$.

Proposition C established that any chosen $\gamma > 0$ is indeed the solution $\gamma^* = \gamma$ of the Lagrangian dual problem for some power $P < \infty$. The following two propositions establish that two different power constraints $P_1 \neq P_2$ correspond to two different shadow prices $\gamma_1^* \neq \gamma_2^*$.

**Proposition D** If $P_1 \neq P_2$, then $\gamma_1^* \neq \gamma_2^*$.

**proof:** Let $P_1 \neq P_2$. By the monotonicity and continuity of $C^{\text{fb}(n)}(P)$, we have

$$C^{\text{fb}(n)}(P_1) \neq C^{\text{fb}(n)}(P_2).$$

Now, assume that the respective solutions to the Lagrangian dual problems are equal, i.e., $\gamma_1^* = \gamma_2^*$. Then

$$C^{\text{fb}(n)}(P_1) = G(\gamma_1^*) = G(\gamma_2^*) = C^{\text{fb}(n)}(P_2),$$

which contradicts our earlier conclusion that $C^{\text{fb}(n)}(P_1) \neq C^{\text{fb}(n)}(P_2)$. Hence, $\gamma_1^* \neq \gamma_2^*$ must hold.

**Proposition E** If $\gamma_1^* < \gamma_2^*$, then $P_1 > P_2$.

**proof:** We observe that for any $\gamma_1^* < \gamma_2^*$ and for any $\mathcal{P}$ (such that $\text{P}_{\text{ow}}(\mathcal{P}) > 0$), we have the following inequality

$$\mathcal{L}(\mathcal{P}, \gamma_1^*) > \mathcal{L}(\mathcal{P}, \gamma_2^*).$$

We first show that

$$G(\gamma_1^*) = \max_{\mathcal{P}} \mathcal{L}(\mathcal{P}, \gamma_1^*) > \max_{\mathcal{P}} \mathcal{L}(\mathcal{P}, \gamma_2^*) = G(\gamma_2^*).$$

Let's assume that the contrary is true, that is

$$\max_{\mathcal{P}} \mathcal{L}(\mathcal{P}, \gamma_1^*) \leq \max_{\mathcal{P}} \mathcal{L}(\mathcal{P}, \gamma_2^*).$$

Let $\mathcal{P}_2^*$ be

$$\mathcal{P}_2^* = \arg\max_{\mathcal{P}} \mathcal{L}(\mathcal{P}, \gamma_2^*).$$

Then we have

$$\mathcal{L}(\mathcal{P}_2^*, \gamma_1^*) \leq \max_{\mathcal{P}} \mathcal{L}(\mathcal{P}, \gamma_1^*)$$
$$\leq \max_{\mathcal{P}} \mathcal{L}(\mathcal{P}, \gamma_2^*)$$
$$= \mathcal{L}(\mathcal{P}_2^*, \gamma_2^*),$$

which contradicts the established inequality $\mathcal{L}(\mathcal{P}, \gamma_1^*) > \mathcal{L}(\mathcal{P}, \gamma_2^*)$ for ay $\mathcal{P}$.

As a result, we have

$$C^{\text{fb}(n)}(P_1) = G(\gamma_1^*) > G(\gamma_2^*) = C^{\text{fb}(n)}(P_2).$$

By the monotonicity and continuity of $C^{\text{fb}(n)}(P)$, we further have

$$P_1 > P_2.$$

**Corollary F** The shadow price $\gamma = \gamma^* > 0$ and the power constraint $P < \infty$ are in a 1-to-1 correspondence.

**proof:** *It is a direct consequence of Propositions D and E.*

## ACKNOWLEDGMENT

The authors would like to thank the anonymous referees for their valuable comments and suggestions.

**Shaohua Yang** (S'99–M'04) received the B.Eng. degree in Electronics Engineering from Tsinghua University, Beijing, China in 2000, and the S.M. and Ph.D. degrees in Electrical Engineering from Harvard University, Cambridge, MA in 2002 and 2004, respectively. From 2004 to 2005, he was a research staff member at Hitachi Global Storage Technologies, San Jose Research Center. He is presently a staff design engineer at Marvell Semiconductor Inc. Dr. Yang's research interests are in information theory, signal processing and coding for applications in communications and magnetic recording.

Dr. Yang received the 2005 IEEE Best Student Paper Award in Signal Processing and Coding for Data Storage.

**Aleksandar Kavčić** (S'93–M'98–SM'04) received the Dipl.-Ing. degree in Electrical Engineering from Ruhr-University, Bochum, Germany in 1993, and the Ph.D. degree in Electrical and Computer Engineering from Carnegie Mellon University, Pittsburgh, Pennsylvania in 1998.

Since 2007 he has been with the University of Hawaii, Honolulu where he is presently Associate Professor of Electrical Engineering. Prior to 2007, he was in the Division of Engineering and Applied Sciences at Harvard University, as Assistant Professor of Electrical Engineering from 1998 to 2002, and as John L. Loeb Associate Professor of Natural Sciences from 2002 to 2006. While on leave from Harvard University, he served as Visiting Associate Professor at the City University of Hong Kong in the Fall of 2005 and as Visiting Scholar at the Chinese University of Hong Kong in the Spring of 2006.

Prof. Kavčić received the IBM Partnership Award in 1999 and the NSF CAREER Award in 2000. He is a co-recipient, with X. Ma and N. Varnica, of the 2005 IEEE Best Paper Award in Signal Processing and Coding for Data Storage. He served on the Editorial Board of the IEEE TRANSACTIONS ON INFORMATION THEORY as Associate Editor for Detection and Estimation from 2001 to 2004. Presently, he is the Chair of the Signal Processing for Storage Technical Committee of the IEEE Communications Society.

**Sekhar Tatikonda** (S'92-M'00) received the Ph.D. degree in Electrical Engineering and Computer Science from the Massachusetts Institute of Technology, Cambridge, in 2000. From 2000 to 2002, he was a Postdoctoral Fellow in the Computer Science Department at the University of California, Berkeley. He is currently an Associate Professor of Electrical Engineering at Yale University, New Haven, CT. His research interests include communication theory, information theory, stochastic control, distributed estimation and control, statistical machine learning, and inference.